\documentclass[preprint2]{aastex60}
\usepackage[T1]{fontenc}
\usepackage{microtype}
\usepackage{afterpage,amsmath,amssymb,graphicx,natbib, color}
\usepackage{hyperref}
\usepackage[english]{babel}
\usepackage{blindtext}
\usepackage{array}   % for \newcolumntype macro
\newcommand\plottwoman[3]{\centering \leavevmode
\includegraphics[width=#3\linewidth]{#1} \hfil
\includegraphics[width=#3\linewidth]{#2}}

\newcommand\plotoneman[2]{\centering \leavevmode
\includegraphics[width=#2\linewidth]{#1}}

\newcommand\plotonebig[1]{\centering \leavevmode
\includegraphics[width=.98\linewidth]{#1}}

\newcommand\citeeg[1]{\citep[e.g.,][]{#1}}
\newcommand\lsim{\lesssim}
\newcommand\gsim{\gtrsim}

\newcommand\Npix{\ensuremath{\mathrm{N_{pix}}}}

\newcommand\Nim{\ensuremath{\mathrm{N}_\mathrm{im}}}

\newcommand\FeH{\ensuremath{[\mathrm{Fe}/\mathrm{H}]}}
\newcommand\ZH{\ensuremath{[\mathrm{Z}/\mathrm{H}]}}
\newcommand\logEBV{\ensuremath{\log \mathrm{E}(\mathrm{B}-\mathrm{V})}}
\newcommand\dmod{\ensuremath{\mu_{d}}}

\newcommand\pcmdpy{\texttt{PCMDPy}}
\newcommand\dynesty{\texttt{dynesty}}
\renewcommand\S[1]{\text{Section }\ref{#1}}
\newcommand\F[1]{\text{Figure }\ref{#1}}
\newcommand\T[1]{\text{Table }\ref{#1}}

\newcommand\HST{\textit{HST}}

\newcommand\changed[1]{{\bf #1}}
\renewcommand\changed[1]{ #1}

\begin{document}

\title{Measuring Star Formation Histories, Distances, and Metallicities with Pixel Color-Magnitude Diagrams II: Applications to Nearby Elliptical Galaxies}

%% Use these options if using aastex62
% \author{B.~A.~Cook}
% \affiliation{Center for Astrophysics | Harvard \&{} Smithsonian, 60 Garden St., Cambridge, MA 02138, USA}
% \correspondingauthor{B.~A.~Cook}
% \email{bcook@cfa.harvard.edu}

% \author{C.~Conroy}
% \affiliation{Center for Astrophysics | Harvard \&{} Smithsonian, 60 Garden St., Cambridge, MA 02138, USA}

% \author{P.~van Dokkum}
% \affiliation{Astronomy Department, Yale University, 52 Hillhouse Ave, New Haven, CT 06511, USA}

%% Use these options if using aastex v6.0
\author{B.~A.~Cook\altaffilmark{1}, Charlie Conroy\altaffilmark{1}, and Pieter van Dokkum\altaffilmark{2}}
\altaffiltext{1}{Center for Astrophysics | Harvard \&{} Smithsonian, 60 Garden St., Cambridge, MA 02138, USA}
\altaffiltext{2}{Astronomy Department, Yale University, 52 Hillhouse Ave, New Haven, CT 06511, USA}
\email{bacook17@gmail.com}

\begin{abstract}

We present spatially-resolved measurements of star formation histories (SFHs), metallicities, and distances in three nearby elliptical galaxies and the bulge of M31 derived using the pixel color-magnitude diagram (pCMD) technique. 
We compute pCMDs from archival \HST{} photometry of M87, M49, NGC 3377 and M31, and fit the data using the new code \pcmdpy{}.
We measure distances to each system that are accurate to $\sim 10\%$.
The recovered non-parametric SFHs place reasonable ($\pm 1$ dex) constraints on the recent (< 2 Gyr) star formation in M31 and NGC 3377, both of which show evidence of inside-out growth.
The SFHs in M87 and M49 are constrained only at the oldest ages.
The pCMD technique is a promising new avenue for studying the evolutionary history of the nearby universe, and is highly complementary to existing stellar population modeling techniques.

\end{abstract}

\section{Introduction}
\label{s.pcmdpy2.intro}

The evolution of galaxies, in particular the build-up of their stellar mass, is shaped by a combination of many physical processes including mergers, feedback from supernovae and active galactic nucleii, and the accretion of gas from the circumgalactic and intergalactic media.
Constraining the impacts of these processes often requires comparisons between observed galaxies and the outputs of hydrodynamical simulations, which can study these physical mechanisms in great detail \citeeg{Hopkins2014,Vogelsberger2014a}.
A galaxy's star formation history (SFH) and chemical enrichment are particularly sensitive to the relative contributions of these processes.

One common method for measuring stellar populations and SFHs in galaxies is resolved-star photometry \citeeg{Dolphin2002,Weisz2011,Lewis2015,Williams2015}, which compares the colors and magnitudes of individual stars to stellar evolution models.
Due to the limited angular resolution of even the best optical telescopes, SFHs measured with the resolved-star photometry technique are primarily limited to galaxies in the Local Group and slightly more distant dwarf galaxies, due to crowding.
Because young, massive main-sequence stars can be identified even against a crowded background, constraining more recent epochs of star formation with resolved stars is notably easier than measuring ancient SFHs, which requires resolving stars as faint as the oldest main-sequence turnoff for the results to be robust against  systematic uncertainties \citep{Schulz2002,Williams2017}.

An alternative method, spectral energy distribution (SED) modeling, compares broadband photometry or spectra of the integrated light of (typically) the entire galaxy against stellar population synthesis models to recover stellar populations and SFHs \citeeg{Walcher2011, Conroy2013}.
The advent of integral-field units (IFUs) allows for spatially-resolved measurements of nearby galaxies with SED-modeling techniques \citeeg{Smith2018a}.
But SFHs derived from SED-modeling are most sensitive to the levels of ongoing star formation, as the light from old, low mass main-sequence stars is usually overwhelmed by the light of a few rare, evolved, young stars \citep{Papovich2001,Maraston2010,Pforr2012a,Sorba2015,Leja2018c}.
The high signal-to-noise required to precisely measure stellar populations using IFUs also often limits spatially-resolved measurements to the brightest inner regions of galaxies.

These two techniques can be categorized within a common framework by considering the typical number of stars per resolution element, a quantity denoted \Npix{} \citep{VanDokkum2014b,Conroy2016}.
The unresolved measurements used in SED-modeling typically have $\Npix \gsim \mathcal{O}(10^6)$, while robustly measuring SFHs by fully resolving the oldest main-sequence turnoff requires $\Npix \lsim \mathcal{O}(10^{-1})$.

In between these two observational regimes lies the so-called \textit{semi-resolved regime}, defined very roughly as $\mathcal{O}(10^1) \lsim \Npix \lsim \mathcal{O}(10^6)$.
All but the brightest individual stars cannot be detected in semi-resolved photometry, but surface-brightness fluctuations from Poisson sampling of rare, bright stars in each pixel are significant. 
Massive galaxies viewed with the Hubble Space Telescope (\HST{}) are typically semi-resolved from $\sim 1$ Mpc out to nearly 100 Mpc. 

A method for measuring stellar populations and SFHs in this regime, known as the pixel color-magnitude diagram (pCMD) technique, was first introduced in \citet{Conroy2016} as a way to model semi-resolved photometry.
The authors demonstrated that the properties of pixel-to-pixel surface-brightness fluctuations are sensitive to the underlying stellar populations and SFH, and made a first application of the method to the bulge of M31.
\changed{Several other works have also studied stellar populations through the distribution of pixels in CMD space, including either qualitative comparisons of pCMDs between galaxies \citep{Bothun1986, Lanyon-Foster2007, Lee2017} or quantifying the dispersion of the pCMD \citep{Lee2018}. However, these works focus on the internal variation of stellar populations across pixels, rather than using the surface-brightness fluctuations as a probe of the stellar populations themselves.}

\citet{Cook2019} presented \pcmdpy{}, a new Python package developed to implement the pCMD technique, with additional levels of model complexity and the ability to simultaneously measure distances.
Mock tests of \pcmdpy{} demonstrated the potential for the pCMD technique to measure spatially-resolved star formation histories (SFHs) and abundances in systems to distances of at least 10 Mpc, and possibly as far as 100 Mpc.

The pCMD technique was also shown to be capable of simultaneously measuring distances to galaxies.
As described in \citet{Cook2019}, the pCMD technique can be considered a generalization of the surface-brightness fluctuations (SBF) distance technique \citep{Tonry1988}.
The SBF metric has been applied to elliptical galaxies and large bulges of spiral galaxies \citep{Ferrarese2000}, but absolute calibration is fairly dependent on uncertain, evolved phases of stellar evolution \citep{Tonry2001}, and the effects of varying stellar populations can be significant.

We present here the first application of the pCMD technique to galaxies outside the Local Group, and the first measurement of distances from pCMDs.
In \S{s.pcmdpy2.data}, we describe a procedure for cleaning, reducing, and aligning archival \HST{} images in order to compute pCMDs that can be analyzed with \pcmdpy{}.
In \S{s.pcmdpy2.methods}, we outline the models used to fit these data.
The results are presented in \S{s.pcmdpy2.results}, including the measured star formation histories (\ref{ss.pcmdpy2.SFHs}), distances (\ref{ss.pcmdpy2.distances}), metal abundance and dust extinction (\ref{ss.pcmdpy2.metallicity}).
We discuss future considerations for the pCMD technique in \S{s.pcmdpy2.discussion}, and conclude in \S{s.pcmdpy2.conclusion}.

\section{\HST{} Data Reduction}\label{s.pcmdpy2.data}

\subsection{Data Collection}\label{ss.pcmdpy2.collection}
We compute pCMDs using archival photometry from the Hubble Space Telescope's Advanced Camera for Surveys (\HST{}-ACS).
We choose three elliptical galaxies with multiple exposures in at least two wide-band filters, and total exposure time in each filter of at least 700 seconds.
The galaxies studied are M87, M49, and NGC 3377. We also study five regions in the bulge of M31, using pre-drizzled data from Brick 01 of the PHAT survey \citep{Dalcanton2012}.
We adopt distance modulus (\dmod) values from the literature, as shown in \T{t.pcmdpy2.distances}, which are used to compute radial distances within a galaxy, and for comparison to our measured distances. The archival data collected for each galaxy are summarized in \T{t.pcmdpy2.data}.

In most galaxies, we use photometry in only two filters: a red ($\mathrm{I}_{814}$ or $\mathrm{z}_{850}$) and a blue ($\mathrm{g}_{475}$) filter.
The exception is M87, where $\mathrm{V}_{606}$ photometry with significantly longer exposure time than the $\mathrm{g}_{475}$ data is also available.
The M87 results use the $\mathrm{I}_{814}-\mathrm{g}_{475}$ color, but we find identical results, within the uncertainties, when using the $\mathrm{V}_{606}$ photometry in place of $\mathrm{g}_{475}$.

\begin{table*}
\centering
\begin{tabular}{|lllcclc|}
\hline
Galaxy & \dmod{} (mag) & D (Mpc) & Method & \changed{Dist. Reference} & \changed{\ZH} & \changed{\ZH{} Reference}\\\hline\hline
M31 & 24.44 & 0.77 & TRGB & \citet{Conn2016} & \changed{0.23} & \changed{\citet{Conroy2012}}\\
NGC 3377 & 30.2 & 10.9 & TRGB & \citet{Lee2016b} & \changed{0.13} & \changed{\citet{Conroy2012}}\\
M87 & 30.9 & 15.1 & TRGB & \citet{Lee2016a} & \changed{0.17} & \changed{\citet{Conroy2012}}\\
M49 & 31.1 & 16.8 & SBF & \citet{Blakeslee2009} & \changed{0.30} & \changed{\citet{VanDokkum2014b}}\\
% NGC 4993 & 33.0 & 40.7 & SBF & \citep{Cantiello2018}\\
%M51 & 29.57 & 8.59 & TRGB & \citep{McQuinn2016}\\
\hline
\end{tabular}
\caption{Benchmark literature distances to the galaxies \changed{and measured metallicities}. We adopt TRGB distances when available, then use distances derived from the SBF technique. We assume these are the true distances, enabling an assessment of the distance measurements with the pCMD technique. Radial distances within a galaxy are determined assuming the distances given here.
\changed{The metallicities listed include variable abundance ratios and computed at various radii, but should be roughly comparable to those measured in this work.}}
\label{t.pcmdpy2.distances}
\end{table*}

\begin{table*}
    \centering
    \begin{tabular}{|lccccc|}
    \hline
    Galaxy & Filter & Exposures & Exp.~Time (s) & Proposal ID & Sky (counts) \\\hline\hline
    M31 & & & & & \\
    & F814W (I) & $\cdots$ & 3430 & 12058 & 201 \\
    & F475W (g)& $\cdots$ & 3800 & 12058 & 114 \\
    NGC 3377 & & & & & \\
    & F850LP (z) & 4 & 3005 & 10554 & 127\\
    & F475W (g) & 4 & 1380 & 10554 & 77\\
    M87 & & & & & \\
    & F814W (I) & 8 & 2880 & 10543 & 293\\
    & F606W (V) & 6 & 3000 & 10543 & 377\\
    & F475W (g) & 2 & 750 & 9401 & 43\\
    M49 & & & & & \\
    & F850LP (z) & 2 & 1120 & 9401 & 57\\
    & F475W (g) & 2 & 750 & 9401 & 50\\
    % NGC 4993 & &&&& \\
    % & F850LP (z) & 2 & 680 & 15329 & 175 \\
    % & F475W (g) & 3 & 1395 & 15329 & 62\\
    %M51 &&&&&\\
    %& F814W (I) & 4 & 1120 & 10452 & 73 \\
    %& F435W (B) & 4 & 2720 & 10452 & 41\\
    \hline
    \end{tabular}
    \caption{Summary of photometric data collected from \HST{} archives. We use drizzled mosaics from the PHAT survey for M31, and so do not reduce individual exposures. The final column lists the estimated sky background, computed using the ACS Exposure Time Calculator, and is in units of counts per pixel, integrated over the entire exposure (see \S{ss.pcmdpy2.background}).}
    \label{t.pcmdpy2.data}
\end{table*}

\subsection{Combining and Aligning Exposures}\label{ss.pcmdpy2.align}

We align each set of individual \texttt{.flc} exposures using \texttt{TweakReg} from the \texttt{AstroDrizzle} package, and then combine exposures with the drizzling tool \citep{Fruchter2010, Gonzaga2012}. 
This increases the effective exposure time of the photometry, removes cosmic rays, and corrects for non-linear optical distortions. 
The \texttt{TinyTim} package \citep{Krist2011} does not provide point-spread function (PSF) models that account for drizzling, so we use the \texttt{lanczos3} kernel, which most faithfully preserves the properties of the original PSF \citep{Jee2007}.
We disable sky subtraction, choosing instead to forward model the addition of sky background during the simulation step (see \S{ss.pcmdpy2.background}).
Regardless, several of our sources are sufficiently extended that the default \texttt{AstroDrizzle} sky subtraction algorithm would significantly overestimate the background.
The defaults are used for all other \texttt{AstroDrizzle} parameters.

Modeling of pCMDs assumes that the images in each filter are extremely well aligned. Yet after the initial drizzling procedure, we find that offsets of as much as a few pixels can remain between filters.
We therefore realign and re-drizzle the final images to a common reference frame, following Section 7.5 of \citet{Gonzaga2012}.
After this second alignment, the images of each galaxy in separate filters are found to be aligned to better than 1/10$^\mathrm{th}$ of a pixel, according to \texttt{TweakReg}.

\subsection{Data Cleaning and Source Removal}\label{ss.pcmdpy2.clean}

The mock tests in \citet{Cook2019} showed that improperly estimating the exposure time can significantly bias the resulting fits, because the effects of photon noise on the pCMD distribution are challenging to disentangle from the true surface-brightness fluctuation signal\footnote{To account for this, the SBF distance technique isolates the true signal by analyzing the fluctuations in Fourier space.}.
We therefore mask any pixels that have an effective exposure time (as computed by \texttt{AstroDrizzle}) lower than $90\%$ of the total in either band.
This removes pixels that were not contributed to by all individual exposures. Common instances include pixels affected by cosmic rays, hot pixels, or regions where the exposures did not overlap, such as across the ACS chip gap.
This results in masking approximately $30\%$ of all pixels.
% We note that handling the contamination from cosmic rays and exposure overlap could present a non-trivial challenge towards stacking arbitrarily large sets of images to reach extreme exposure times.

We mask all detectable sources in the images (globular clusters, foreground stars, and background galaxies) which will have significantly different properties in pCMD space than the diffuse field star populations in the target galaxy.
We first use Source Extractor \citep{Bertin1996, Barbary2016} to subtract a smoothed model of the galaxy and then automatically identify and mask the majority of sources.
We then mask by-eye any residual sources, such as obvious globular clusters near galaxy centers, and stray light from extended background galaxies and bright foreground stars. 

\subsection{pCMD Extraction}\label{ss.pcmdpy2.extraction}

The final step in extracting pCMDs from data is to identify regions where the standard assumption of modeling pCMDs -- that the typical number of stars in a pixel (\Npix{}) is roughly constant -- is at least approximately valid.
To this end, we focus on elliptical galaxies (and the bulge of M31) in this work, as they typically show far less substructure than spiral galaxies. 
Yet all galaxies exhibit significant radial surface-brightness gradients, making it impractical to apply the pCMD technique to the entire galaxy.

To overcome this limitation, we extract pCMDs from thin, roughly elliptical regions in each galaxy, within which the average surface-brightness does not vary significantly.
This yields the added benefit of allowing us to measure spatially-resolved SFHs and metallicities as a function of radius.
In selecting such regions, however, there is a balancing act required between selecting slices that are too wide (such that surface-brightness gradients distort the pCMD distribution) and those that are too thin (such that too few pixels remain to be analyzed).
For this reason, we are unable to extract valid pCMDs in the central regions of each galaxy (roughly within 10"), and instead focus on the outer regions, as described below.

We compute elliptical slices aligned with the galaxies using SAOImage DS9 \citep{Joye2003}, by creating highly smoothed contours (smoothness $\approx 120$) and converting them to polygonal regions.
The contours are scaled logarithmically, and spaced such that the typical surface-brightness gradient across each region is significantly less than 0.1 magnitudes.
In practice, this results in regions of roughly 20 pixels in width.
In each galaxy, we select three regions (five in M31), equally spaced in radius, and further divide these into four quadrants to minimize \changed{the effects of} any large-scale azimuthal asymmetry.
We primarily focus our analysis on one quadrant from each radius, although we confirmed that the results below are consistent for each quadrant.

\changed{Any remaining large-scale variations in surface-brightness (either due to azimuthal variations or overly-wide annuli) in the extracted regions will result in broadened pCMD distributions.
Mock tests indicate that the primary impact on the inferred parameters is an underestimate of \Npix{} (and hence distance) to match the scale of surface-brightness fluctuations, although the magnitude of the bias is challenging to estimate and depends significantly on \Npix{}.}

We compute pCMDs by first converting the instrumental fluxes to apparent magnitudes, using zero-points computed using the \texttt{pysynphot} package \citep{Lim2015}.
We extract separate pCMDs from each of the regions, and remove the masked pixels as described above.
Example pCMDs from three regions of M87 are shown in \F{f.pcmdpy2.M87pcmds}.

% \begin{figure*}
% \plotoneman{M87_labeled}{0.95}
% \caption{The regions studied in M87.
% Shown in grayscale is the archival I-band \HST{} photometry used in this study.
% The red regions show the roughly-elliptical contours from which we extract the pCMDs (the thin slices in between each pair of contours). Three radial regions, labeled as A, B, and C, are studied, and each is divided into 4 quadrants, shown by the dashed lines.
% Sources (globular clusters, background galaxies, and foreground stars) are masked prior to extracting the pCMDs.}
% \label{f.pcmdpy2.M87}
% \end{figure*}

\begin{figure}
\plotoneman{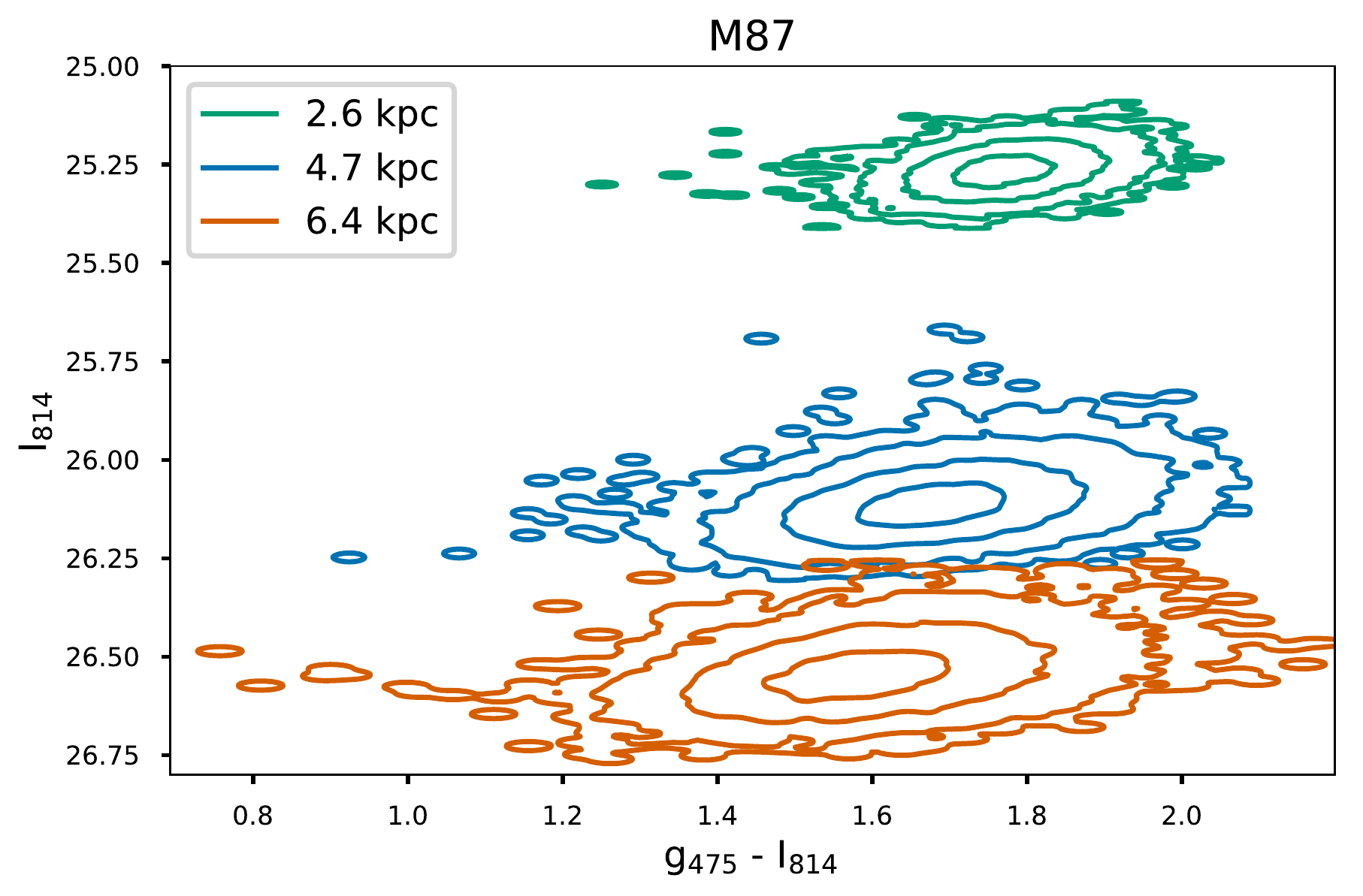}{0.98}
\caption{Example pCMDs extracted from quadrants of the M87 photometry, at the three radii given. Contours show the bounds within which 39\%, 87\%, 99\%, and 99.9\% of the pixels lie (the $1\sigma$, $2\sigma$, $3\sigma$, and $4\sigma$ contours, respectively).
The decrease in average flux and increase in variance with radius demonstrates the effect of decreasing \Npix.}
\label{f.pcmdpy2.M87pcmds}
\end{figure}

\subsection{Background Estimation}\label{ss.pcmdpy2.background}

Background (sky) noise has a significant impact on the observed pCMDs, and must be considered carefully.
\changed{If we consider sky noise to be a constant background in each filter, then its primary effects are to 1) increase the average flux in each pixel, and 2) change the level of photon noise measured at the CCD.
Accurately modeling photon noise is crucial to modeling pCMDs, because it can easily be mistaken for intrinsic surface-brightness fluctuations \citep{Cook2019}. Therefore, we add a synthetic sky signal (a constant count-per-pixel rate in each filter) into our simulated pCMDs, rather than subtracting the sky background from our photometry.}

Several of our sources, especially M87 and M49, are widely extended on the sky, such that the entire \HST{} field of view is contaminated by emission from their diffuse envelopes.
This makes automated methods for estimating the sky, such as those provided in \texttt{AstroDrizzle}, impractical, as the background would be overestimated by including emission from the source.

We instead estimate the sky background using \HST{}'s online Exposure Time Calculator (ETC)\footnote{\href{http://etc.stsci.edu/etc/input/acs/imaging/}{http://etc.stsci.edu/etc/input/acs/imaging/}}.
Given the observed filters, location of the source, and date of observation, ETC computes the expected background count rate from zodiacal light, which we convert to counts-per-pixel integrated over the entire exposure.
The sky signal \changed{we add into each simulated filter} are listed in \T{t.pcmdpy2.data}.
These estimates also include the background contribution from Earthshine, using the models provided by the ETC and available pointing data from each observation. In most cases the Earthshine background is very small compared to the zodiacal light.
We validate our background estimates in NGC 3377, which is far less extended than M87 and M49, by computing the median flux in the corners of the image farthest from the source. The values agree with our estimates to within $5\%$.
\changed{The diffuse extragalactic background is expected to be less than a $10\%$ additional contribution \citep{Zackrisson2009}, and is not included.}

Noise from dark current can also have a significant effect on the level of photon noise, in some regions contributing as much as 10$\%$ of the total flux.
Dark current is already subtracted from the \texttt{.flc} exposures, and we mask hot and warm pixels in the data cleaning stage above.
To account for the average dark current of the remaining pixels, we add the ETC estimated dark current (0.0127 counts per pixel per second) to our simulated images prior to applying Poisson photon noise, and then subtract the same values from the results.

\section{Methods}\label{s.pcmdpy2.methods}

\subsection{Overview of \pcmdpy{}}
\label{ss.pcmdpy2.pcmdpy}

We analyze the data using \pcmdpy{}, a Python code developed to fit pCMDs and infer the Bayesian posterior over physical parameters.
We provide here a brief overview of the code, while full details can be found in \citet{Cook2019}.

Pixel Color-Magnitude Diagrams with Python (\pcmdpy{}) is a GPU-accelerated python code for inferring physical parameters of semi-resolved galaxies through fitting observed pCMDs to synthetic models.
It relies on the key assumption of pCMD modeling, that the stars in each pixel are drawn via a Poisson process from the same underlying population, with an average number of stars per pixel given by a parameter \Npix{}.
The pCMDs simulated by \pcmdpy{} therefore are only applicable to small regions of a galaxy where this assumption is valid.

Simulating a pCMD with \pcmdpy{} begins with an assumed model for the metal abundances and SFH (with the total stars formed summing to \Npix) of a region, and an initial mass function \citep[by default,][]{Salpeter1955}.
Stars from this model are randomly populated into each simulated pixel of an image, and the flux in each pixel is computed using the MIST stellar evolution models \citep{Choi2016}.
We apply dust attenuation, adjust for the modeled distance to the source, and finally include observational noise such as sky background, PSF convolution, and photon noise according to the properties of the telescope and the simulated exposure time.
The fluxes in each pixel are converted to magnitudes, resulting in a pixel color-magnitude diagram.

The primary free parameters of the physical model specify the metal abundance, star formation history, dust attenuation, and distance, with multiple options for each model available in \pcmdpy{}.
The likelihood, or agreement between data and modelled pCMDs, is computed by binning the pixels into a Hess diagram and comparing the relative counts.
The posterior distribution is estimated using the nested sampling code \dynesty{} \citep{Speagle2019}.

\subsection{\pcmdpy{} models}
\label{ss.pcmdpy2.models}

Three model changes have been implemented relative to that described in \citet{Cook2019}.
The first, as described in \S{ss.pcmdpy2.background}, is the addition of dark current noise, which is added prior to simulating the Poisson photon noise and then subtracted.

Secondly, we have removed the sub-pixel PSF model, and instead apply a single PSF to the entire image, because subsequent tests indicate the particular sub-pixel PSF model described in \citet{Conroy2016} and \citet{Cook2019} is insufficient for modeling the desired effects. 
We refer the reader to the discussion in \S{ss.pcmdpy2.psf}.

Finally, we have added an additional metallicity distribution function (MDF) model that replicates the MDF of a "closed box" evolution model \citep{Binney1998}.
\changed{The closed box MDF has a similar shape to a normal distribution, but benefits from having only a single free parameter, as the width scales with the mean metallicity. The closed box model may not, of course, be a perfect representation of the shape of the MDF in a narrow elliptical annulus of an ETG, and future work should consider whether better physically-motivated distributions, such as the "leaky-box" model, result in any significant changes to the derived physical properties.}

As our default model, we fit the pCMDs with a 5-bin non-parametric SFH, a single metallicity, and single dust extinction screen, and allow the distance modulus to vary (Models M1+S5+E1+D2, from \citet{Cook2019} Table 1), for a total of 8 free parameters.
We assume flat priors over all parameters in the model, including $\FeH$ (from -1.0 to 0.5), $\logEBV$ (from -2.0 to -0.5), and $\dmod$ ($\pm 2$ magnitudes around the literature distance).
For each galaxy, we visually identify an approximate $\Npix$ through comparisons to simulated pCMDs with a 10 Gyr SSP.
We then assume a flat prior in the 5 star formation history parameters, with limits of $\pm 2$ dex around a $\tau=3$ Gyr model with that approximate $\Npix$. 

Model images have $\Nim=512$ ($512^2$ pixels), and we assume a Salpeter IMF \citep{Salpeter1955}.
Posteriors are fit with the \dynesty{} dynamic nested sampling algorithm \citep{Speagle2019}, using 400 live points and uniform sampling within a multi-ellipsoidal boundary.
Each full model takes around 120 GPU-hours to fit, executed on NVidia Tesla K20xm chips.

In addition to the default model described above, we study two additional models. One assumes a fixed distance, equal to the literature value given in \T{t.pcmdpy2.distances}, while the other replaces the single metallicity assumption with a closed box MDF. 

Throughout this work, parameter estimates and error-bars are reported as the median and $68\%$ credible interval of the marginalized posterior distribution.

\subsection{Likelihood Model and Statistical Convergence}\label{ss.pcmdpy2.convergence}

As discussed in \citet{Cook2019}, we implement a post-processing correction to the final \dynesty{} weights to account for biases arising from the stochastic likelihood function.
Given the very wide priors we assume here, we find the stochastic effects and decreased sampling efficiency to be much more significant than in earlier mock tests.
We found that in most cases, the fits did not statistically converge.
The sampling efficiency declined so significantly that it took several thousand likelihood calls to produce each of the final few hundred sampled points.

Even using the post-processing correction described in \citet{Cook2019} often resulted in only a few representative samples (with non-negligible posterior weights).
We therefore applied a more liberal correction, lowering the log-likelihood ceiling until the nested sampling convergence metric ($\Delta \ln Z$) was less than $0.01$.
This results in weighting more evenly over many more of the final sampled points.
We discuss this in greater detail in \S{ss.pcmdpy2.likelihood}, but note that the errorbars presented here may be overestimated.

\section{Results}\label{s.pcmdpy2.results}

\subsection{Overview}\label{ss.pcmdpy2.results_overview}
\begin{figure*}
\plotonebig{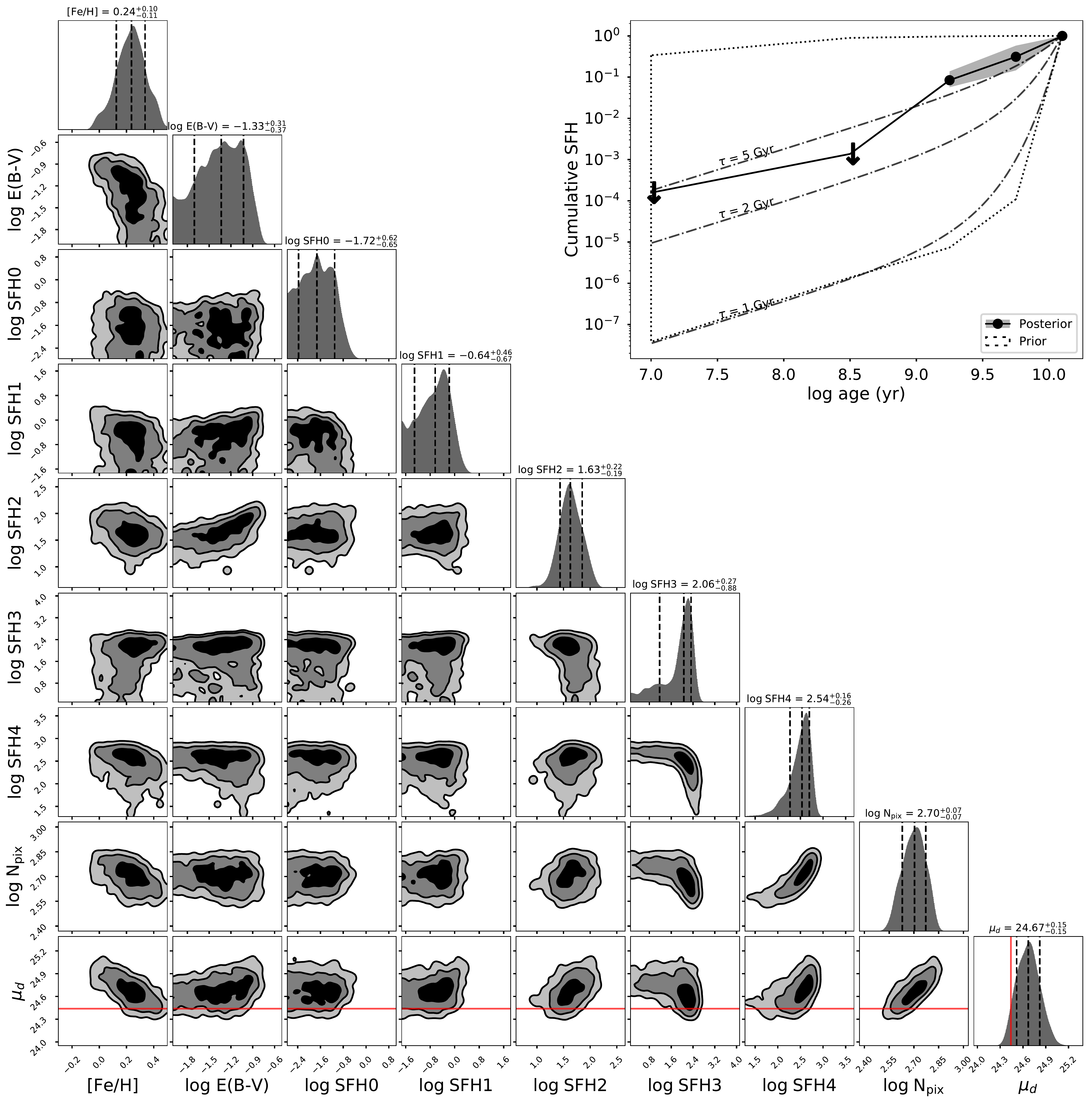}
\caption{Recovered posterior probability distribution for M31 Region E, at a radius of 1.2 kpc. The $1\sigma$, $2\sigma$, and $3\sigma$ contours are shown.
The literature distance modulus is shown as a red line.
The SFH parameters have units of stars-per-pixel, and \Npix{} is computed from their sum. 
The upper-right panel shows the distribution of cumulative star formation history ($68\%$ credible interval).
The allowed prior region is shown as a dotted box, and dot-dashed lines show comparisons to $\tau$-SFHs.
We cannot constrain lower limits to the star formation rates in the youngest two bins, but there is evidence of a decrease in the star formation rate of more than an order of magnitude over the log age interval from 9.25 to 8.5.}
\label{f.pcmdpy2.m31corner}
\end{figure*}

\begin{figure*}
\plotonebig{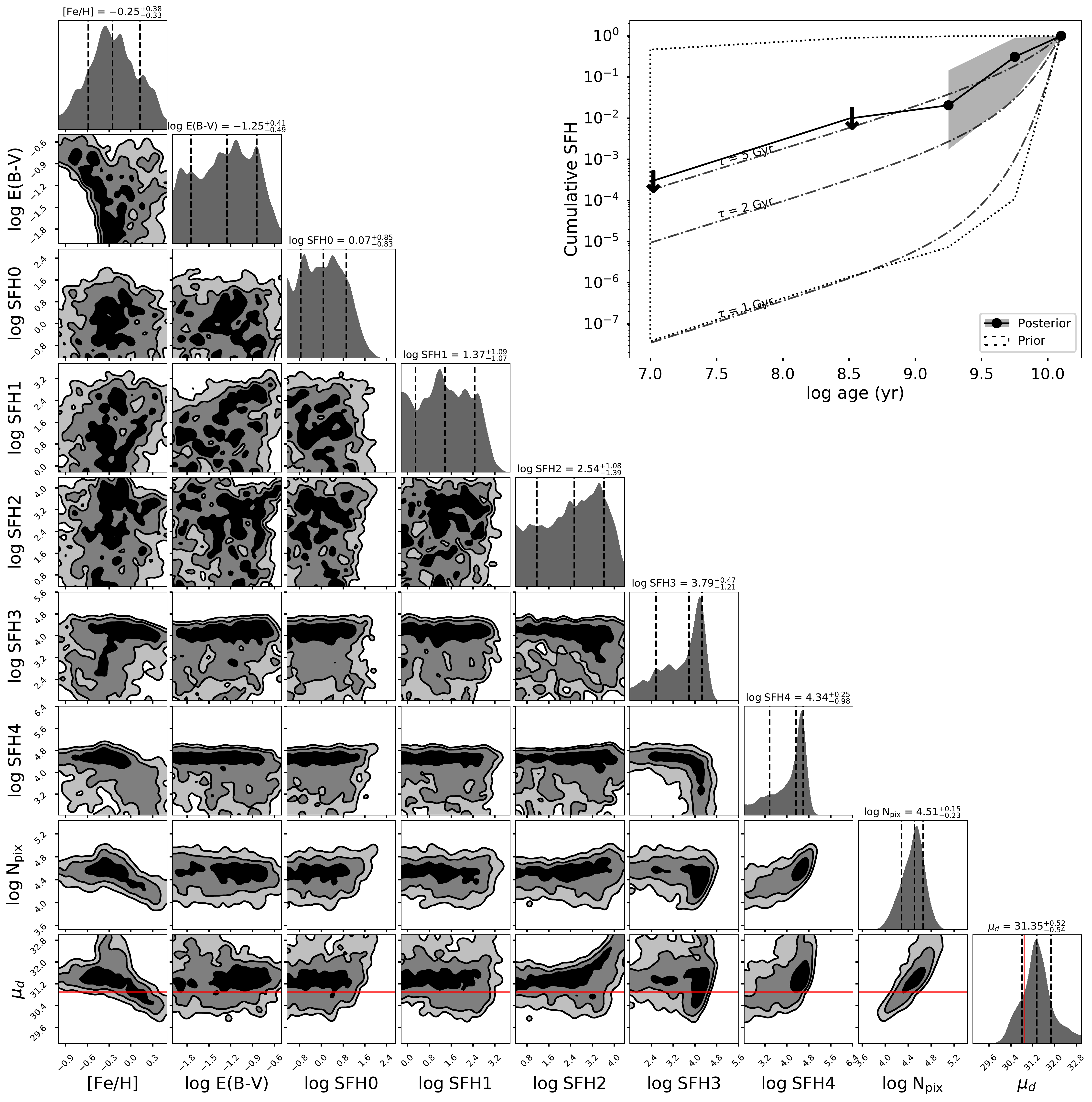}
\caption{Same as \F{f.pcmdpy2.m31corner}, for Region C1 of M87, at a radius of 6 kpc.}
\label{f.pcmdpy2.m87corner}
\end{figure*}

The final output of our modeling procedure are samples representing the 8-dimensional posterior probability distribution. 
A convenient way to visualize this distribution is through a "corner plot", showing the marginalized posterior distribution between all sets of two parameters.
We show examples from Region E (1 kpc) in M31 and Region C1 (6 kpc) in M87, in Figures \ref{f.pcmdpy2.m31corner} and \ref{f.pcmdpy2.m87corner}, respectively.
In each figure, we also show the inferred cumulative SFH, with the $68\%$ confidence region.

In the outer M31 region, we see evidence for most of the same parameter degeneracies as in the mock tests of \citet{Cook2019}. The distance is strongly degenerate with \Npix, and there is also a notable dust-metallicity degeneracy.
There are also strong correlations between the star formation parameters, especially at the two oldest ages (SFH3 and SFH4), but together the overall amount of old star formation is well constrained.

The M31 region shows strong evidence for relatively elevated (more than a $\tau = 5$ Gyr model) levels of star formation through the past $\sim 1$ Gyr, before apparently undergoing a quenching period.
We see evidence of a sharp decrease (by more than an order of magnitude) in the star formation rate between 2 and 0.3 Gyr ago.
The old star formation is less well constrained in M87, but is marginally inconsistent with very rapid and early quenching (such as a $\tau = 1$ Gyr model). \changed{We discuss this further in \S{ss.pcmdpy2.SFHs}.}

In both cases, we are only able to place upper limits to the amount of star formation in the youngest two bins.
The \dynesty{} fits do constrain the upper limits, but the tail of allowed star formation stretches to the lower limits of our flat priors.
Despite the large luminosities and very blue colors of young, massive stars, they are so rare that they will populate only a very small number of pixels, and as such the data are not constraining between very low levels and none at all.
As such, even significantly expanding the prior volume will not result in a converged lower limit.
We show the $68\%$ upper limit in such cases, and discuss this further in \S{ss.pcmdpy2.young}.

\begin{figure*}
\plotonebig{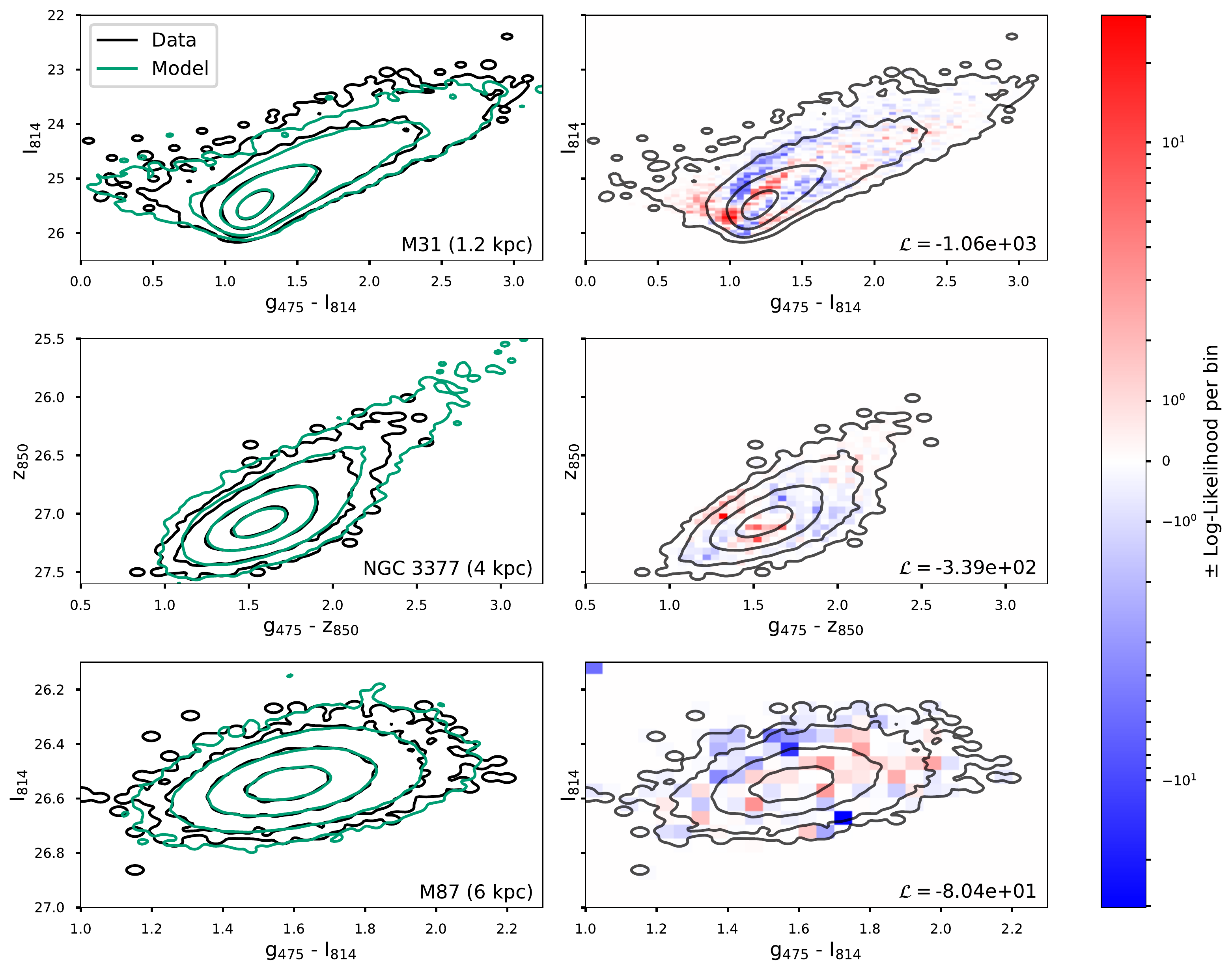}
\caption{Comparisons between example datasets and their corresponding best-fit single metallicity models. The \textit{left} column shows the contours of the data (black) and model (green) pCMDs as in \F{f.pcmdpy2.M87pcmds}. The \textit{right} column shows the likelihood grid as a heatmap in Hess diagram space, multiplied by $\pm 1$ to show over- (under-)densities of model points in red (blue).
The data pCMD is overplotted for reference.
The total log-likelihood of the model is given in the bottom-right corner.
The pCMDs and residuals for M49 are qualitatively similar to M87, and so are not shown.
The pCMD contours and residuals for the closed box MDF models are also visually similar to the single metallicity models.}
\label{f.pcmdpy2.residuals}
\end{figure*}

The models are able to replicate well the input data in each of the galaxies studied, and examples are demonstrated in the residual plots shown in \F{f.pcmdpy2.residuals}.
The pCMD distributions of data and best-fit models are overlaid, showing a high degree of similarity. 

\subsection{Spatially-Resolved Star Formation Histories}\label{ss.pcmdpy2.SFHs}

\begin{figure*}
\plottwoman{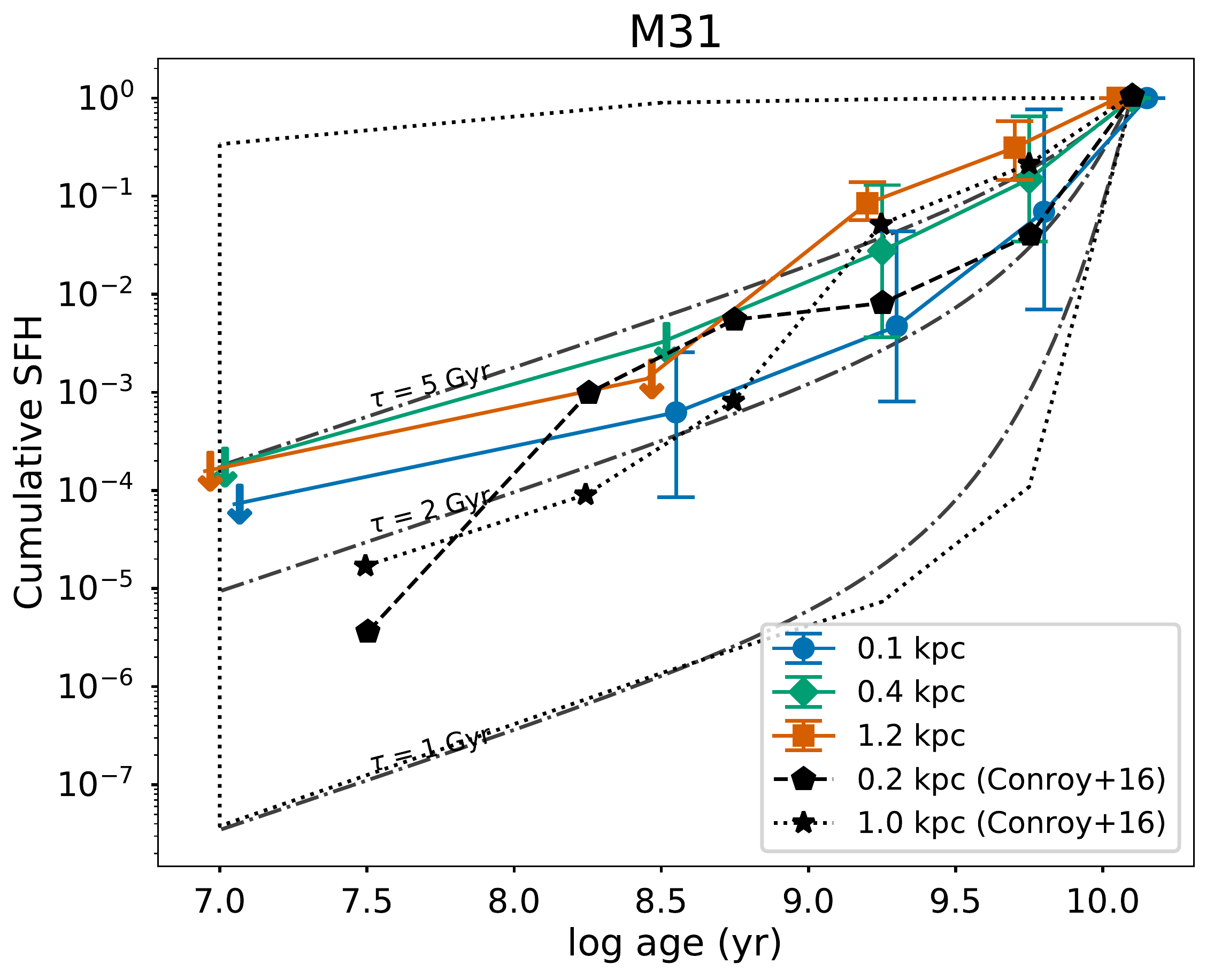}{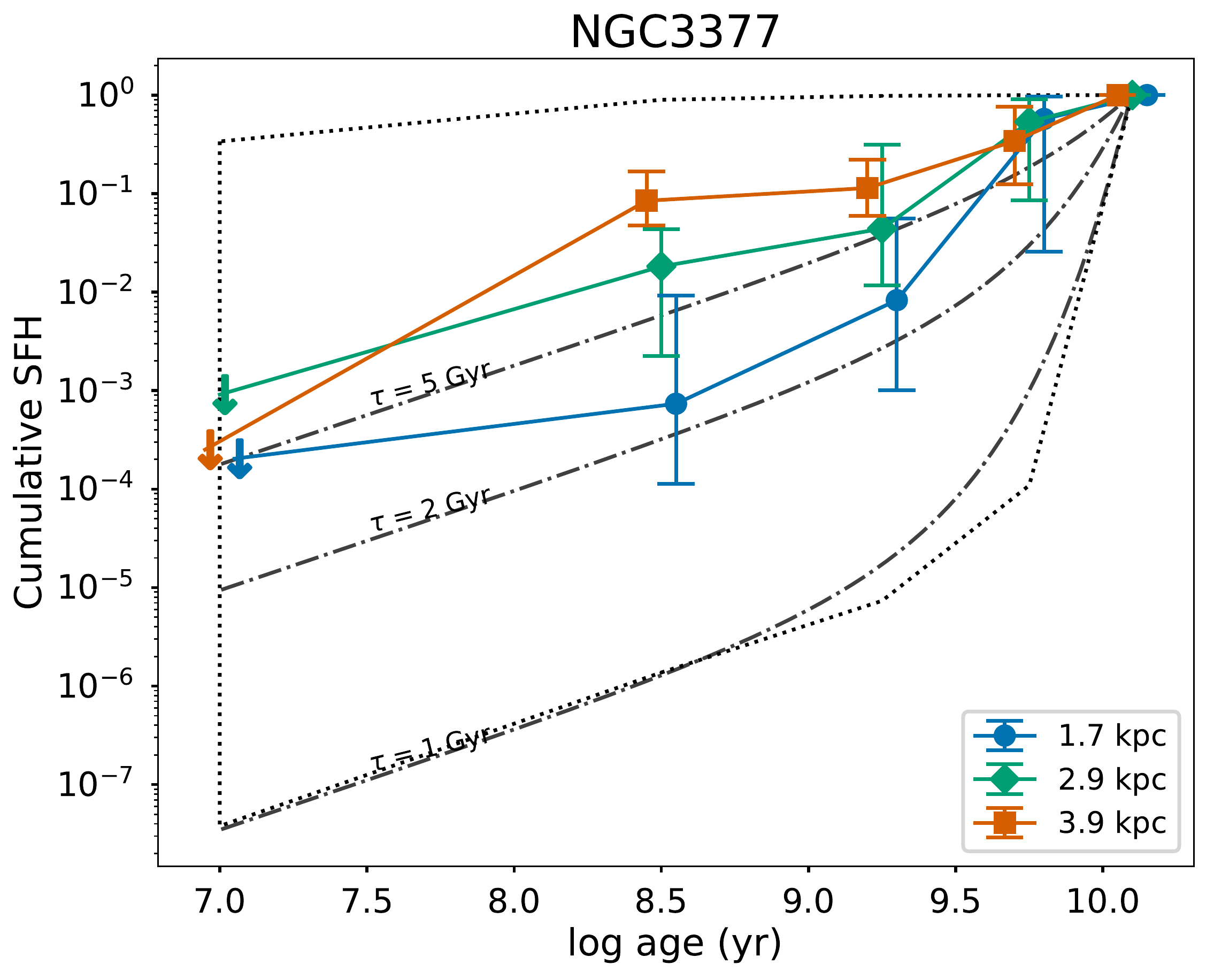}{0.49}
\plottwoman{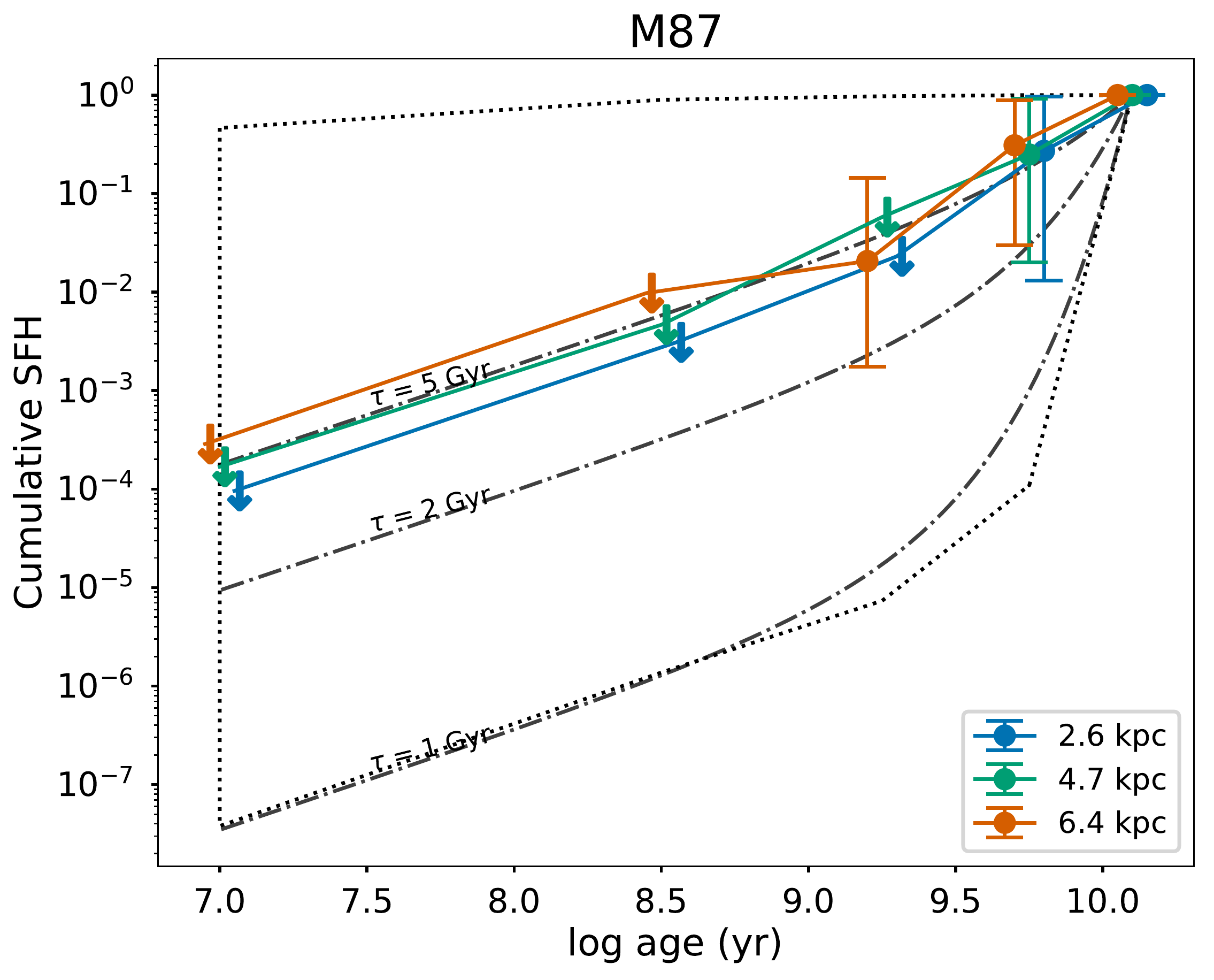}{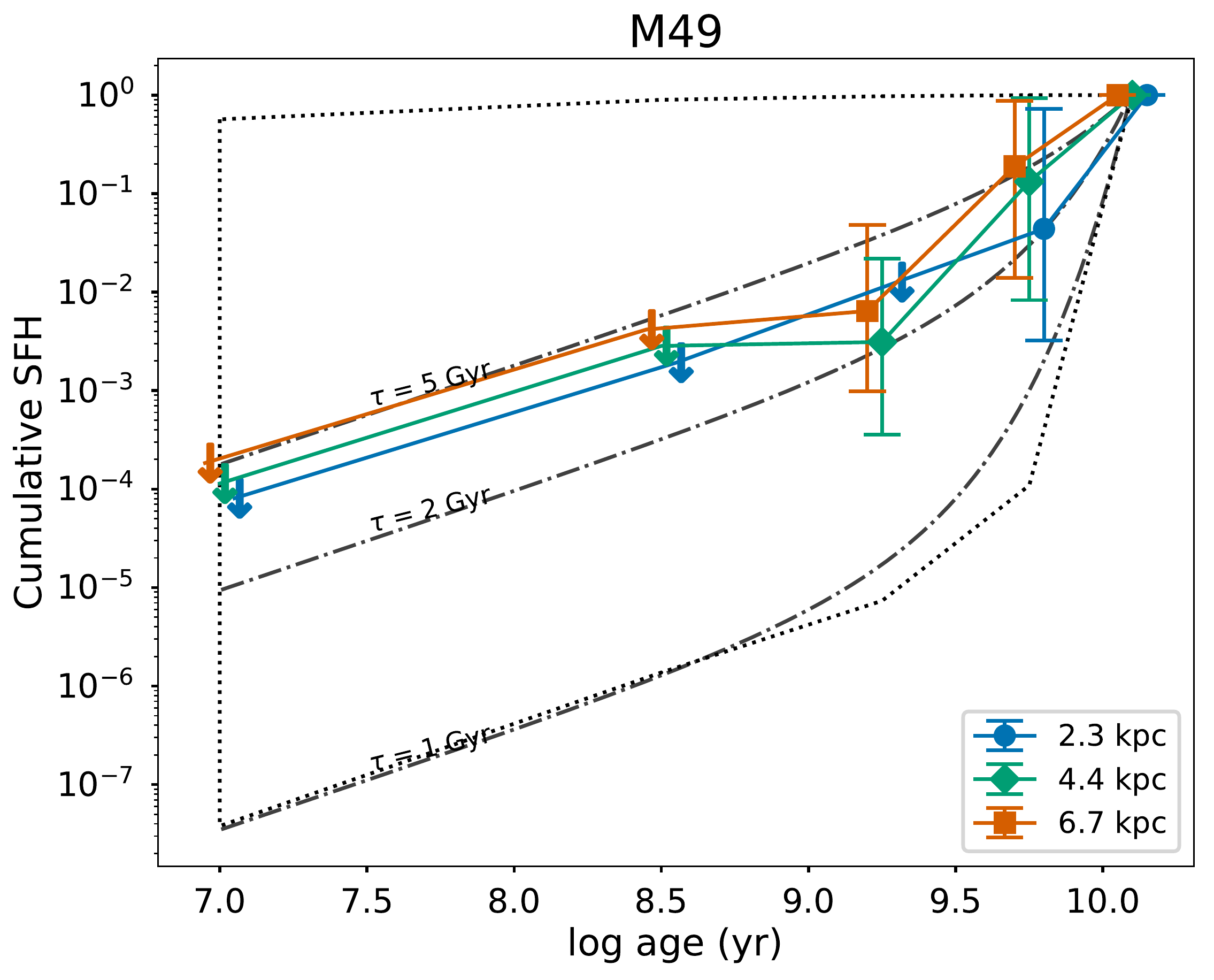}{0.49}  
\caption{The cumulative SFH (with $68\%$ intervals) for three regions in the M31 bulge (\textit{upper-left}), NGC 3377 (\textit{upper-right}), M87 (\textit{lower-left}), and M49 (\textit{lower-right}). Prior boundary, upper limits, and example $\tau$ SFHs as in \F{f.pcmdpy2.m31corner}.
For clarity, we plot points with small offsets in age and replace shaded $68\%$ intervals with error bars.
In the \textit{upper-left} panel, we show comparisons with \citet{Conroy2016}, who first applied the pCMD technique to derive SFHs in M31, and our measurements agree within the uncertainties, although we find the measurements of young star formation to be only upper limits.
Both M31 and NGC 3377 show evidence for longer periods of star formation in the outer regions, whereas the inner regions likely quenched earlier.}
\label{f.pcmdpy2.sfh}
\end{figure*}

\begin{figure}
\plotonebig{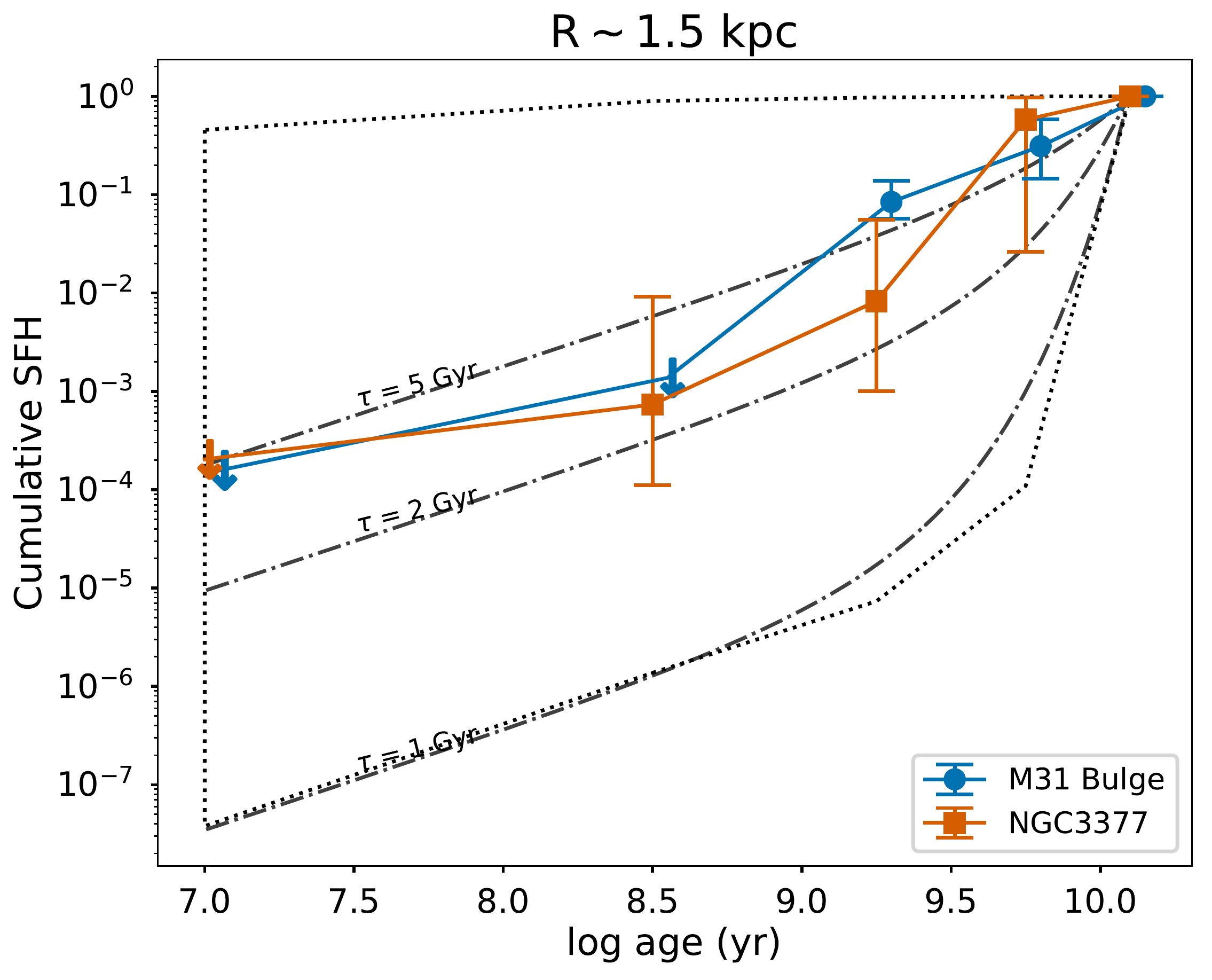}
\plotonebig{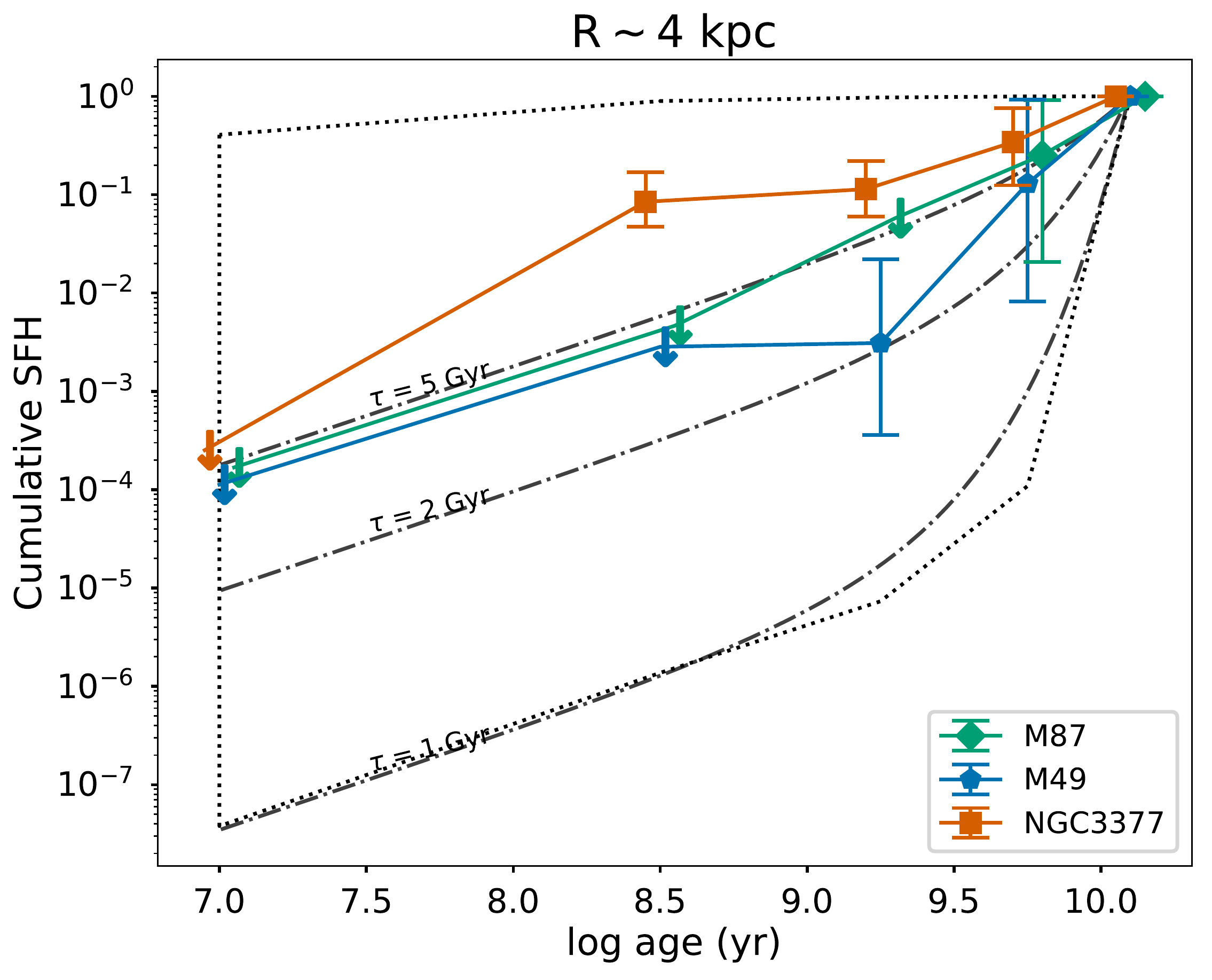}
\caption{Comparison of the cumulative SFHs of galaxies at similar physical radii.
\textit{Top:} Regions approximately 1.5 kpc from the center of M31 (1.2 kpc) and NGC 3377 (1.7 kpc).
\textit{Bottom:} Regions around 4 kpc from the centers of NGC3377 (3.9 kpc), M87 (4.7 kpc) and M49 (4.4 kpc).}
\label{f.pcmdpy2.sfh_compare}
\end{figure}

\begin{figure}
\plotonebig{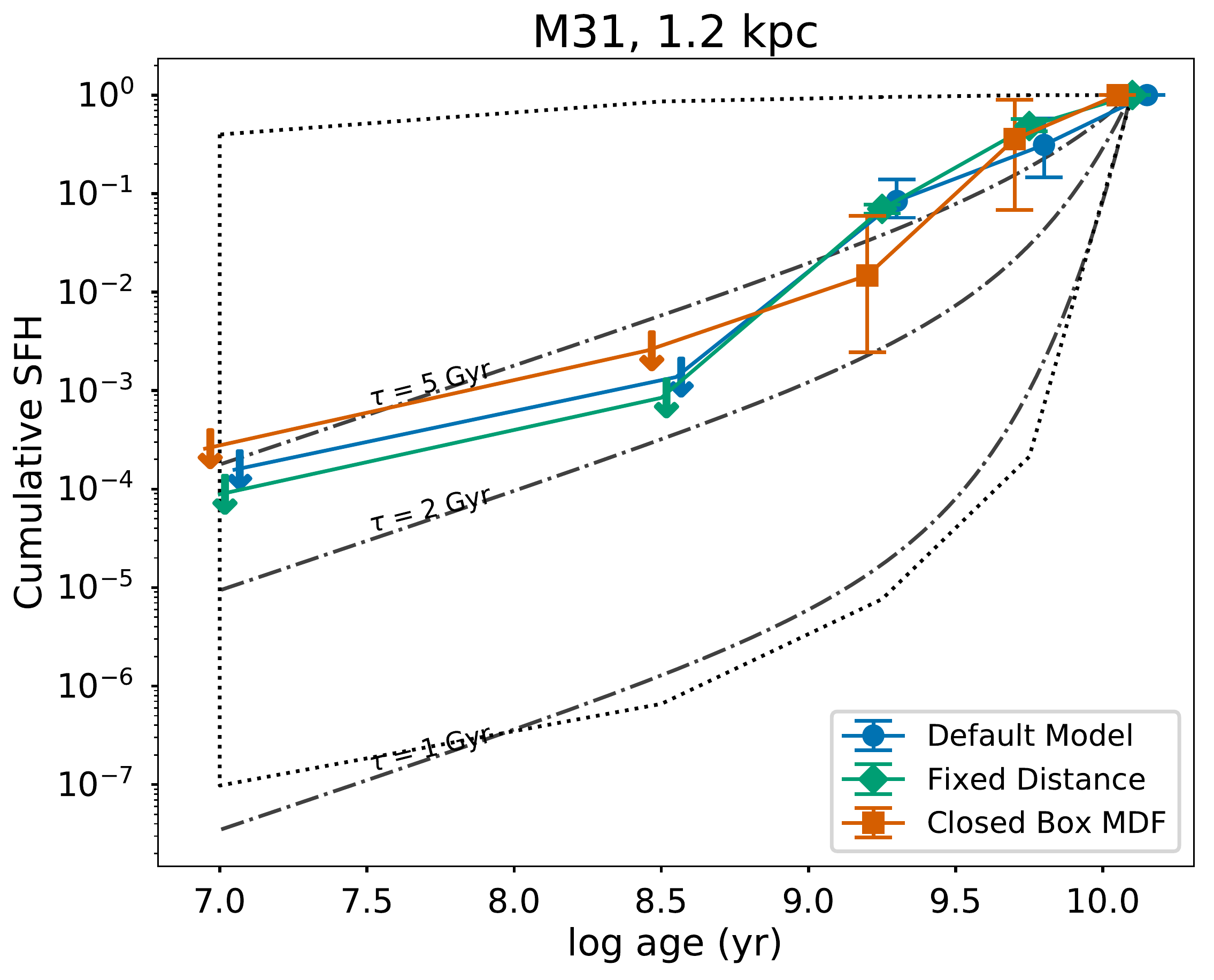}
\plotonebig{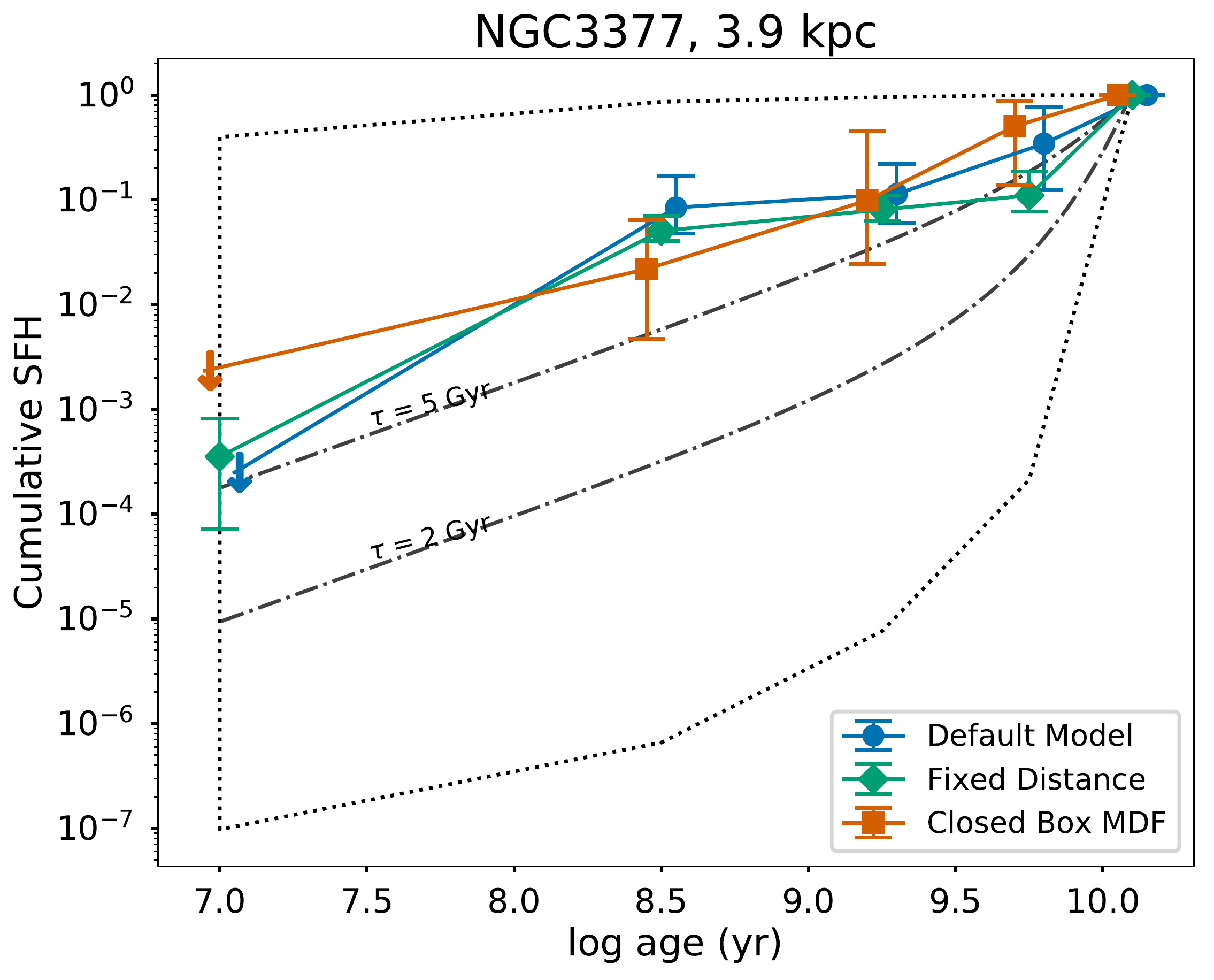}
\caption{The cumulative SFHs as a function of model assumed.
We show the outer regions studied in the M31 bulge (\textit{upper}) and NGC 3377 (\textit{lower}), the only two datasets with noticeable differences between the measured SFHs.}
\label{f.pcmdpy2.sfh_models}
\end{figure}

Among the primary results from the pCMD modeling are measurements of the spatially-resolved star formation history in each of the galaxies.
The posterior distributions of cumulative star formation, as a function of radius, are shown for the four galaxies in \F{f.pcmdpy2.sfh}.

\citet{Conroy2016} previously applied the pCMD technique to derive the cumulative SFH in M31, and we show comparisons to their results in \F{f.pcmdpy2.sfh}.
At the oldest ages, our measured SFHs agree well with their results, as we find evidence for increased levels of star formation in the outer regions, and evidence of earlier quenching in the inner $\sim 0.2$ kpc.
While the oldest bins of star formation agree within the uncertainties with \citet{Conroy2016}, we emphasize that we believe the only an upper limit on the youngest periods of star formation can be constrained.

In NGC 3377, we see evidence for significant gradients in the SFH. The outer region shows strong evidence of elevated star formation, with as much as $10\%$ of the stars formed within around the last 300 Myr \footnote{\changed{although this is sensitive to SFH model, see \ref{f.pcmdpy2.sfh_models}}}. The inner region, by comparison, appears to have quenched much earlier, with significantly less than $1\%$ of its stars forming in the last Gyr.

In contrast to the relatively well constrained histories in M31 and NGC 3377, the SFHs in M87 and M49 are, within the uncertainties, consistent at all radii.
The old ($\gsim 1$ Gyr) star formation in both systems is consistent with $\tau$ models, with exponential timescales between $2-5$ Gyr.
Only upper limits can be placed on the youngest star formation, \changed{but the limits are inconsistent with the levels of recent star formation expected from a $\tau \approx 5$ Gyr model. These findings, that old SFHs are consistent with $\tau\approx5$ Gyr while the recent SFHs are not, may indicate that $\tau$ models are too inflexible to model realistic SFHs \citeeg{Carnall2018b,Leja2019}.}

In \F{f.pcmdpy2.sfh_compare}, we compare SFHs between galaxies, overlaying the posterior distributions at similar radii.
We do not have overlap between all four galaxies at any individual radius.

In the inner $\sim$1 kpc, the bulge of M31 has a more extended history of star formation than NGC 3377, forming around $10\%$ of its stars in the last $\sim1$ Gyr.
Yet at 4 kpc, NGC 3377 has a much younger SFH than M49 and M87.
At all radii, the star formation histories of M49 and M87 are generally consistent, within the uncertainties.

It is important to properly understand the effects of different model assumptions on the inferred stellar populations.
We show, in \F{f.pcmdpy2.sfh_models}, the changes in derived star formation histories in M31 (Region E) and NGC 3377 (Region C1) due to different assumed physical models.
These two regions were the only two with noticeable changes in inferred SFH between models.

When distance is held fixed, the inferred histories of star formation are remarkably consistent, especially given that the distance measured in the NGC 3377 region shown is underestimated by around $30\%$ (see \S{ss.pcmdpy2.distances}).
The only notable offset is the second oldest SFH bin in NGC 3377, where the star-formation decreases slightly.

Assuming a closed box MDF distribution rather than a single metallicity does slightly alter the inferred SFHs, suggesting somewhat earlier quenching in M31 and a marginally smoother SFH in NGC 3377.

In M31 and NGC 3377, the spatially-resolved SFHs indicate a general pattern of more recent star formation with increasing radius.
This is further demonstrated in \F{f.pcmdpy2.younggrads}, where we show the fraction of mass formed within the last 2 Gyr as a function of radius.
Because of the small differences between the inferred SFHs by metallicity model, we show the trends for both models, yet the results are similar in both cases.

Despite the large uncertainties, the outer regions of NGC 3377 and M31 both show evidence for more recent buildup of mass: the outer M31 field formed more than $1\%$ of its stars in the last 2 Gyr, and as much as $10\%$ of the stars in the outer NGC 3377 region formed as recently.
These results are in general agreement with the inside-out formation scenario \citep{White1991,Roskar2008,Rix2013,Patel2013}.
The young star formation in M87 and M49 is too poorly constrained to shown any indication of inside-out growth.

\begin{figure*}
\plotonebig{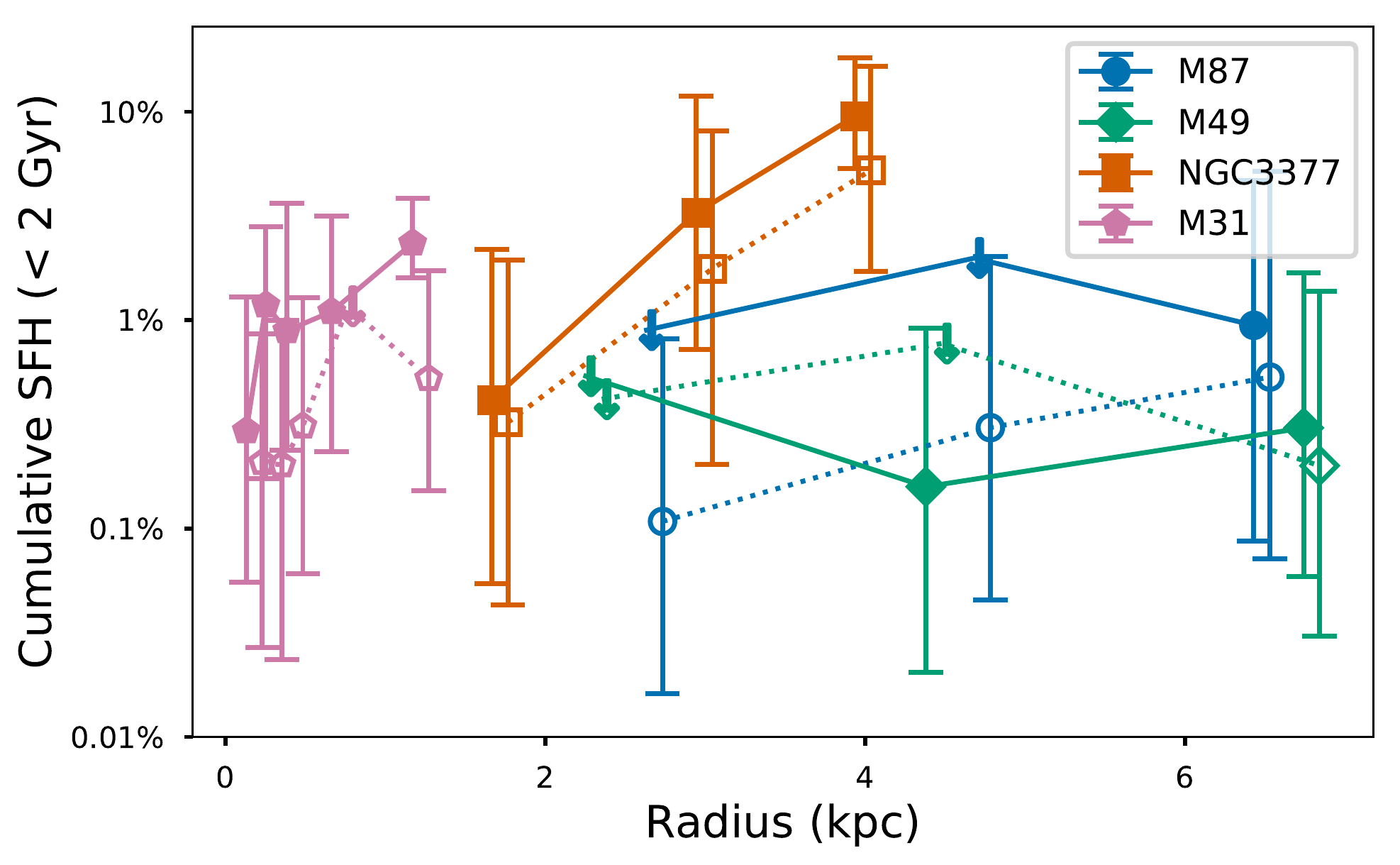}
\caption{Spatially-resolved young (< 2 Gyr) mass fraction.
From each SFH distribution, we derive the fraction of mass formed within last 2 Gyr.
The \textit{solid} symbols show results from the single metallicity model, while \textit{open} symbols show ages from the closed box MDF models.
\changed{Upper-limits are shown where the star-formation history over the past 2 Gyr was not constrained.}
M31 and NGC 3377 show hints of inside-out growth (more recent star formation at larger radius), but this is not constrained in M49 and M87.}
\label{f.pcmdpy2.younggrads}
\end{figure*}

\subsection{Distances with the pCMD Technique}\label{ss.pcmdpy2.distances}

\begin{figure}
\plotonebig{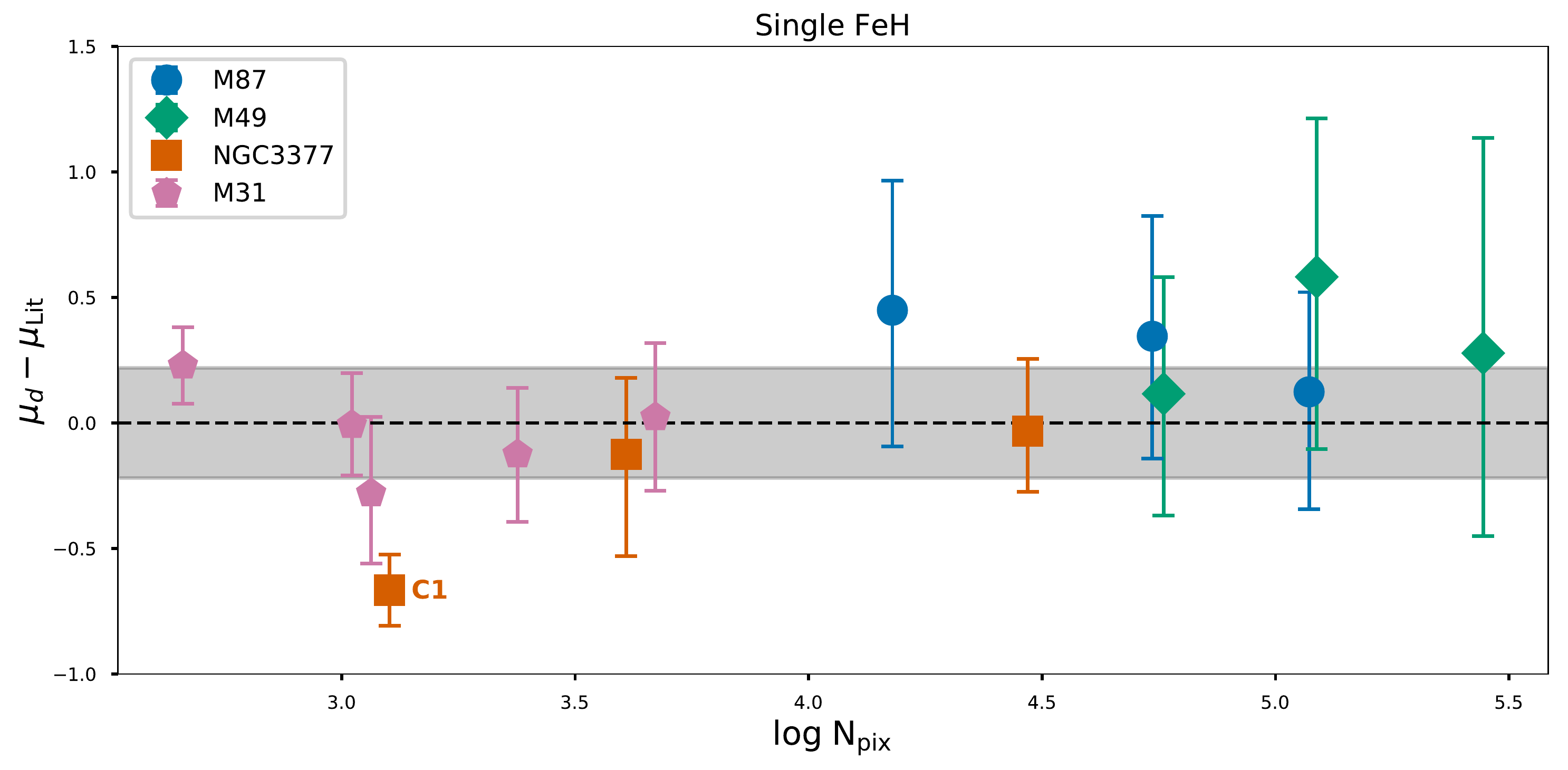}
\plotonebig{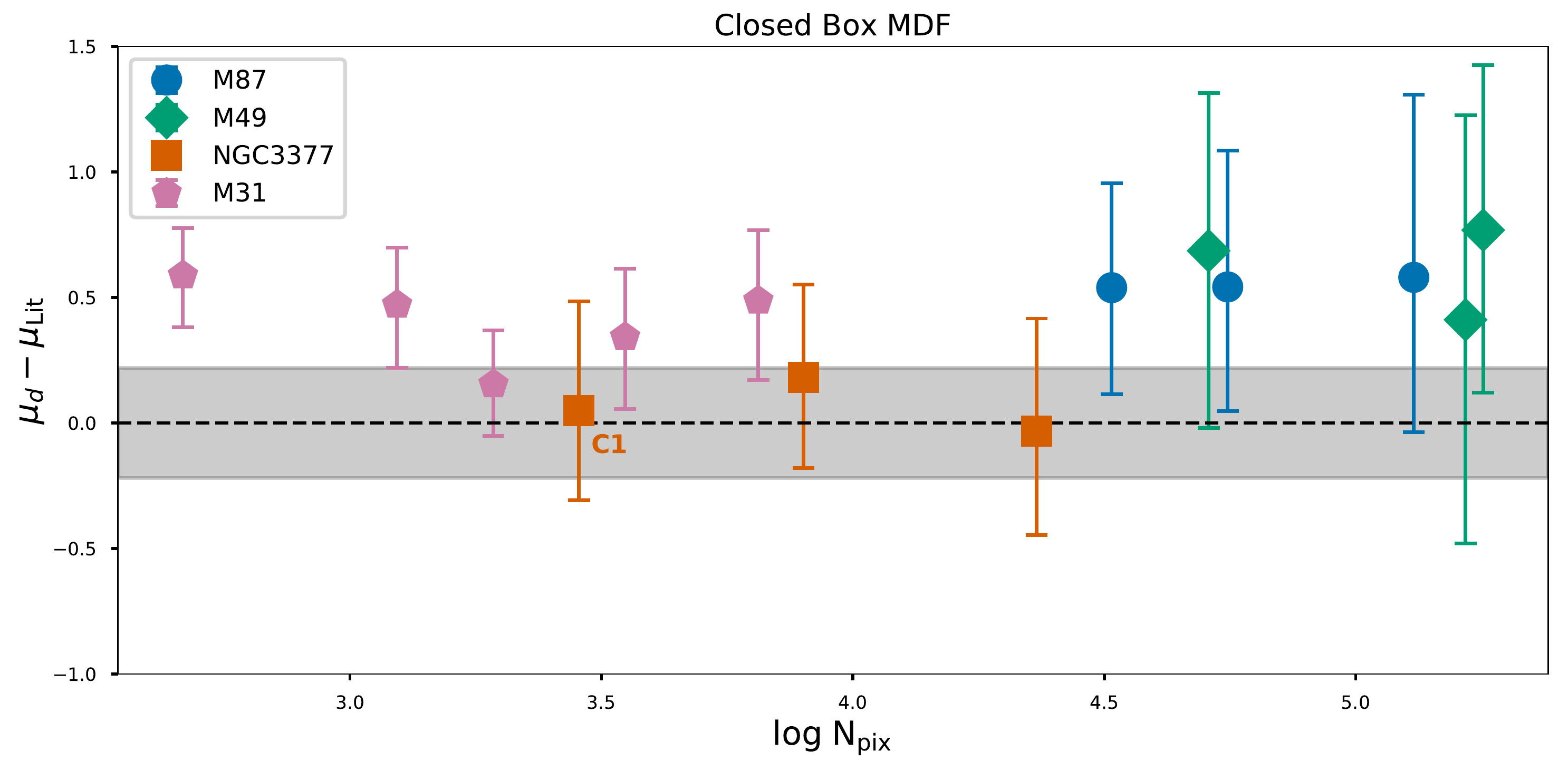}
\caption{The measured distances to each galaxy region, compared to the adopted literature values, derived from the single metallicity (\textit{top}) and closed box MDF (\textit{bottom}) model fits and as a function of \Npix{}.
The grey shaded region shows a $10\%$ distance uncertainty ($\pm 0.22$ magnitudes).
Region C1 in NGC3377 is noted, as its distance measured with the Single FeH model is inconsistent with the literature value.
}
\label{f.pcmdpy2.distances}
\end{figure}

\begin{figure*}
\plottwoman{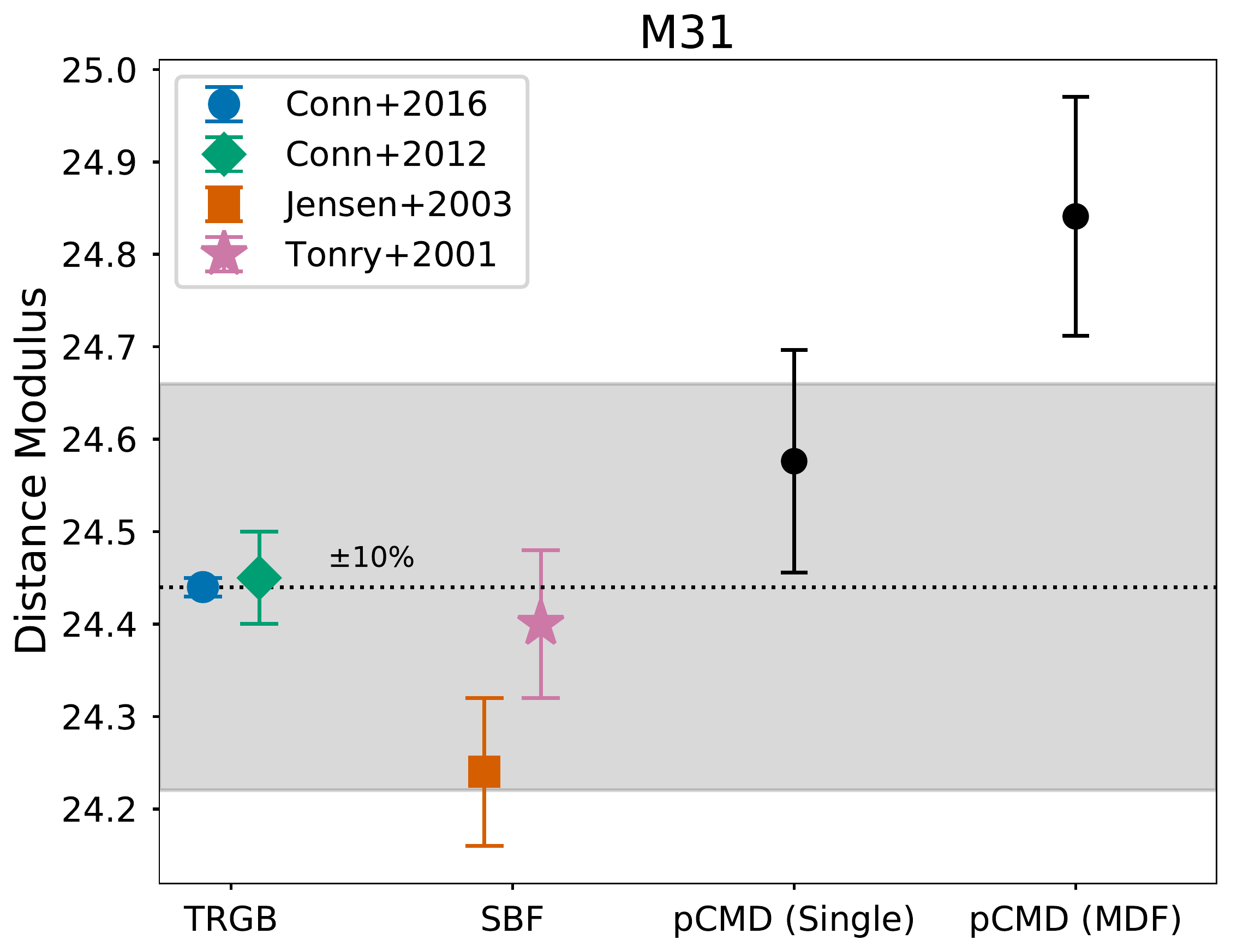}{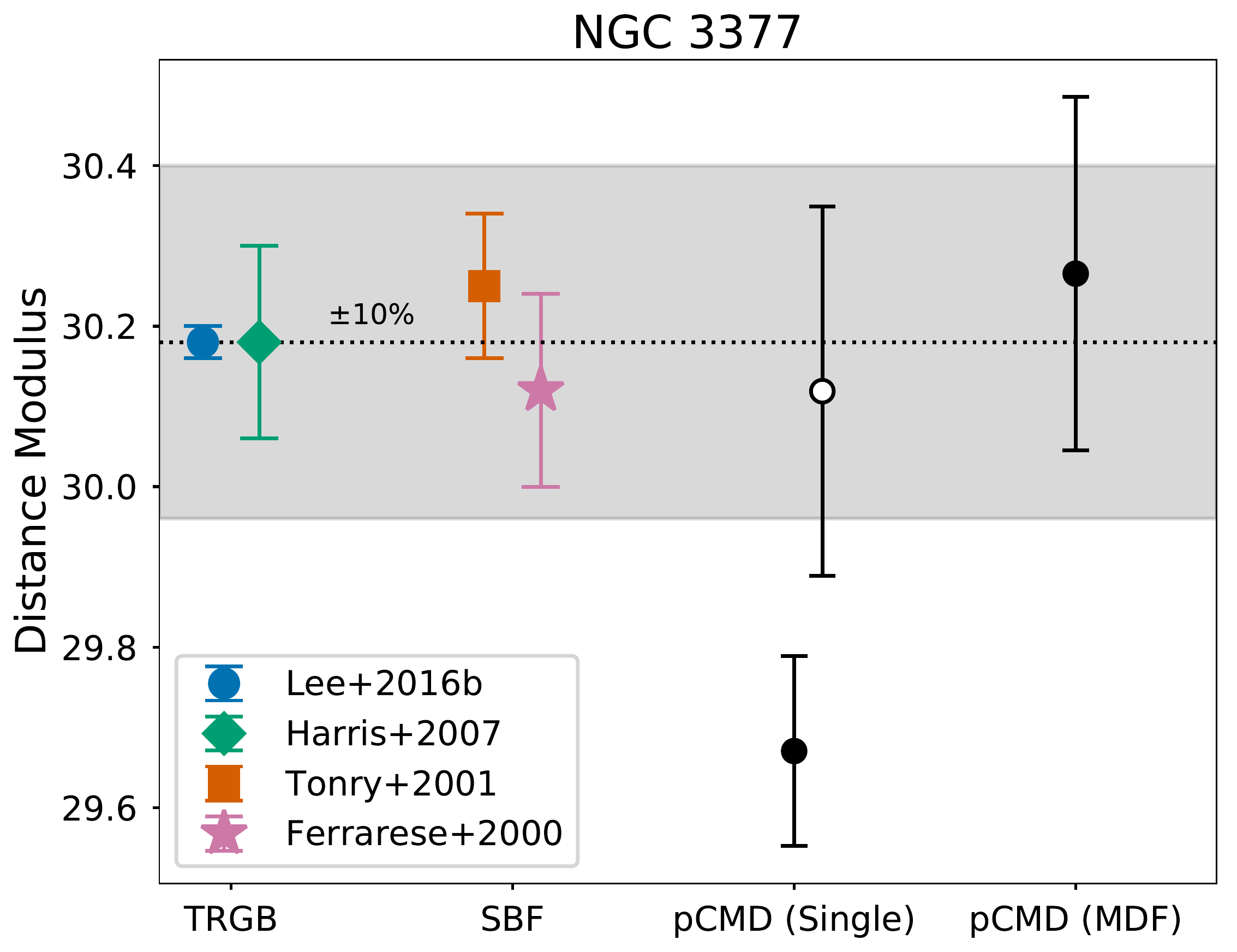}{.49}
\plottwoman{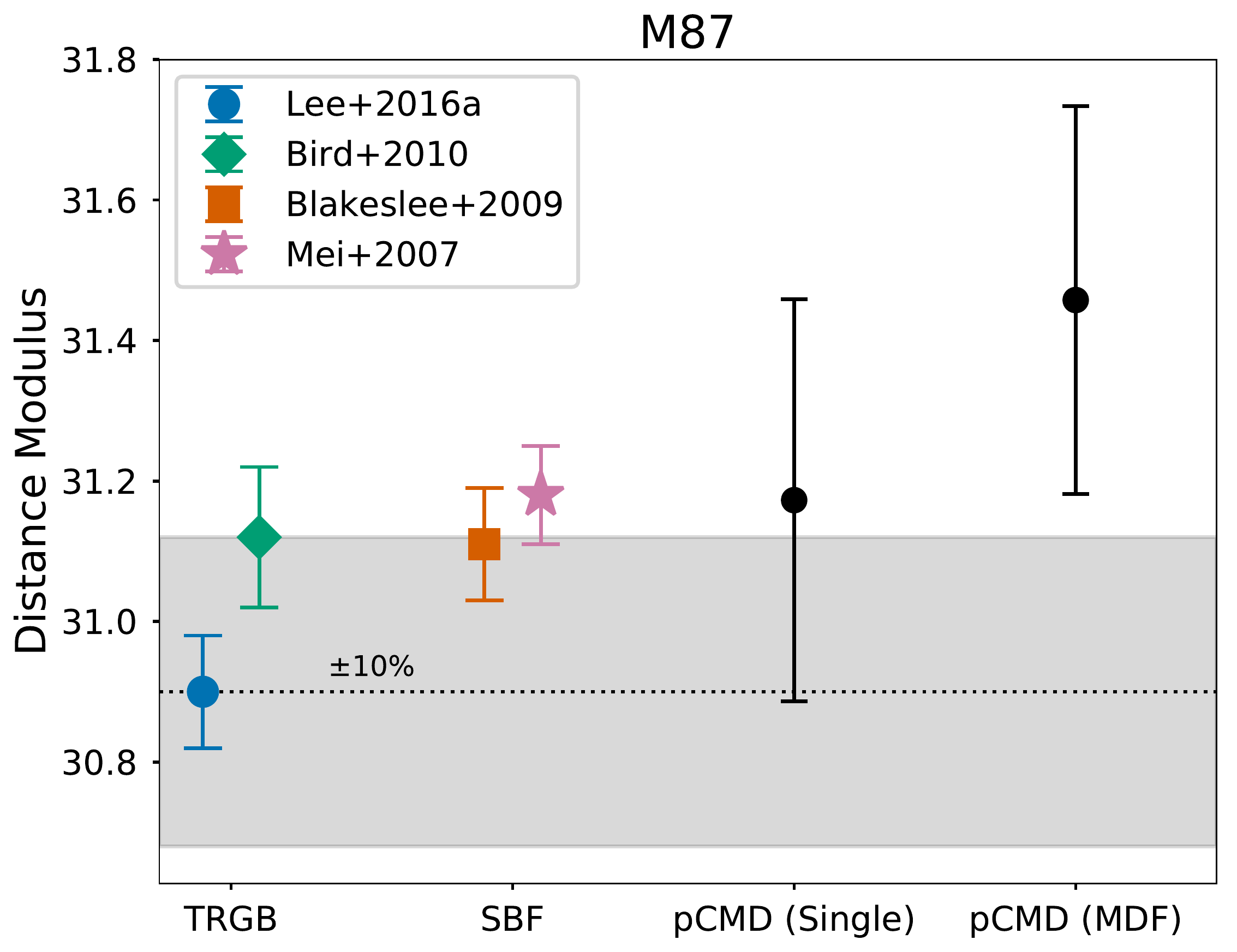}{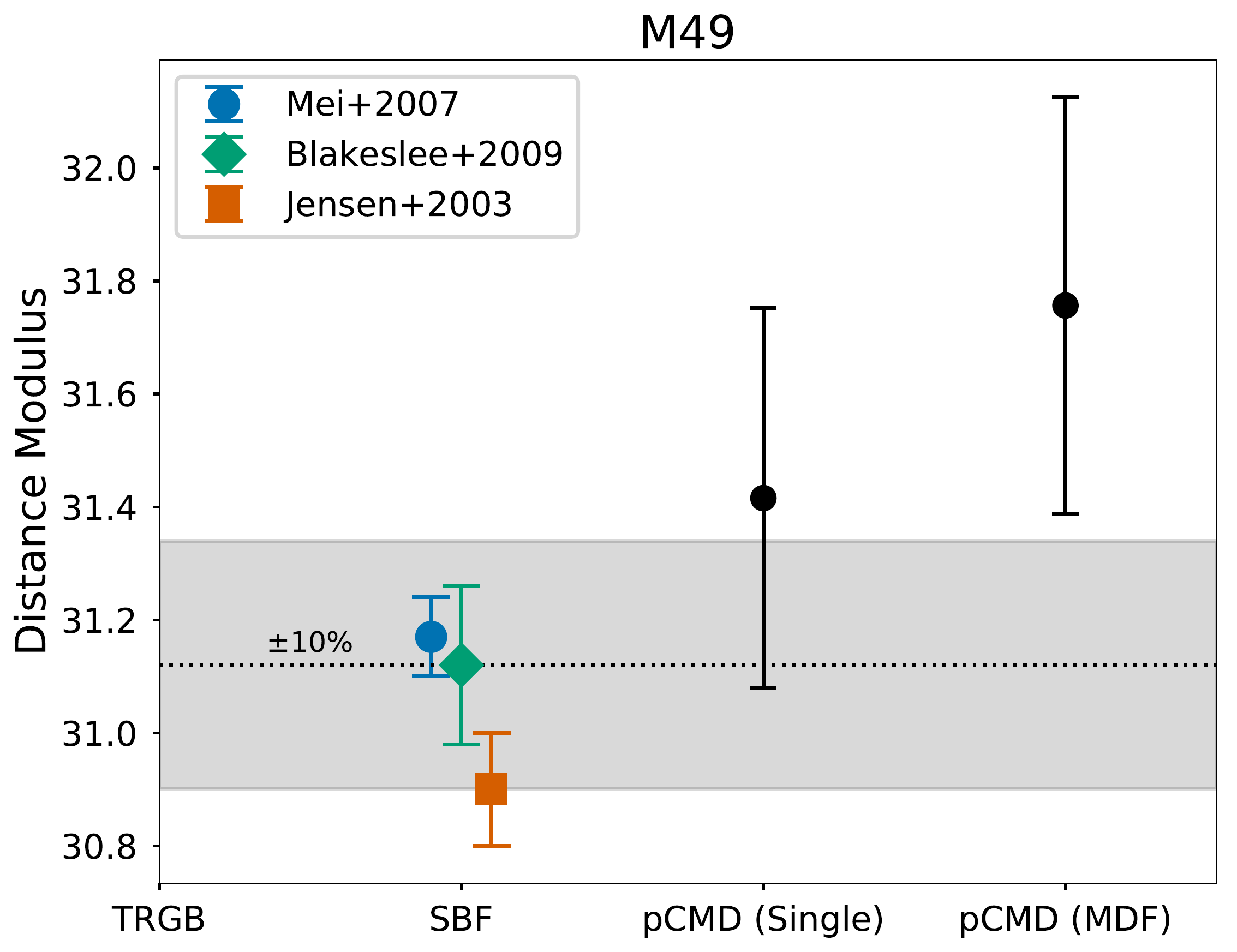}{.49}
\caption{Comparison of distances to each galaxy by method.
Literature distances via the TRGB method and the SBF method are shown, as well as the weighted average of distances from each region in \F{f.pcmdpy2.distances}.
The pCMD distances show estimates from both assumed metallicity models.
For comparison, the gray shaded region shows a $\pm 10\%$ distance uncertainty around the literature value from \T{t.pcmdpy2.distances}.
No TRGB distances to M49 were found in the NED database.
All methods are also subject systematic calibration uncertainties of order $\sim 0.1$ magnitudes (not shown).
The open circle for NGC 3377 shows the effect of removing the apparent outlier, region C1, where we believe the sky background estimate may be contributing to the poor fit.
}
\label{f.pcmdpy2.distance_metrics}
\end{figure*}

In addition to measuring spatially-resolved SFHs, \citet{Cook2019} demonstrated the capability of measuring distances with the pCMD technique.
We recover a separate measurement of the distance modulus to a galaxy from each region.
In \F{f.pcmdpy2.distances}, we show the posterior distribution of these distances, relative to the literature values, for both the single metallicity and closed box MDF models.

In the single metallicity case, we find good agreement to within $\pm 10\%$ in all cases, with the exception of the outer region (C1) of NGC 3377 (see \F{f.pcmdpy2.distances}), as the distance estimate is biased low by around 0.6 magnitudes.
The best-fit metallicity in this region (see \F{f.pcmdpy2.feh_grads}) is also significantly higher than the other regions in NGC 3377.
We suspect this may be caused by a moderate underestimate in the assumed sky background, which the models may compensate for by preferring a closer distance to the source and which would affect the fainter outskirts more significantly.

In the case of the closed box abundance model, the measured distances appear to be systematically biased high by $20\%$ on average, but we do note the distances to NGC 3377 are in better agreement.
The typical distances to M87 and M49 are likewise biased high, but are consistent with the true values within the uncertainties.
We attribute the overestimated distances to the fact that the closed box model is known to overestimate the width of the metallicity distribution functions in observed galaxies \citeeg{Holmberg2007}.

We summarize the distance estimates to each of the four galaxies in 
\F{f.pcmdpy2.distance_metrics}, where we compare the distances derived with the pCMD technique to literature distances measured with the TRGB and SBF techniques.
We take the pCMD method estimate to be the weighted average of the individual measurements from each region.

Both the TRGB distances and the SBF distances are subject to systematic uncertainties of $\sim0.1$ magnitudes (not shown in \F{f.pcmdpy2.distance_metrics}).
In the case of TRGB, this arises from the uncertainty in the TRGB magnitude \citep{Lee2016a}, while for SBF it is due to calibration relative to the Cepheid distance metric \citep{Blakeslee2009}.
The pCMD technique would likely be subject to a similar level of systematic uncertainties as the TRGB, as it relies on well-calibrated luminosities of the brightest stars.

The average distances measured with the pCMD technique agree very well with the existing methods, although again we note the potential systematic differences between the two metallicity models.
Including this possible bias, the distances measured with pCMDs appear reliable to around $10-20\%$, and the pCMD technique is therefore very useful for simultaneously making approximate estimates of distances and stellar populations of galaxies within 100 Mpc.
Improving the precision of measured distances will likely require significant improvements to the likelihood model (see \S{ss.pcmdpy2.likelihood}).

\subsection{Metal Enrichment and Dust}\label{ss.pcmdpy2.metallicity}

\begin{figure}
\plotonebig{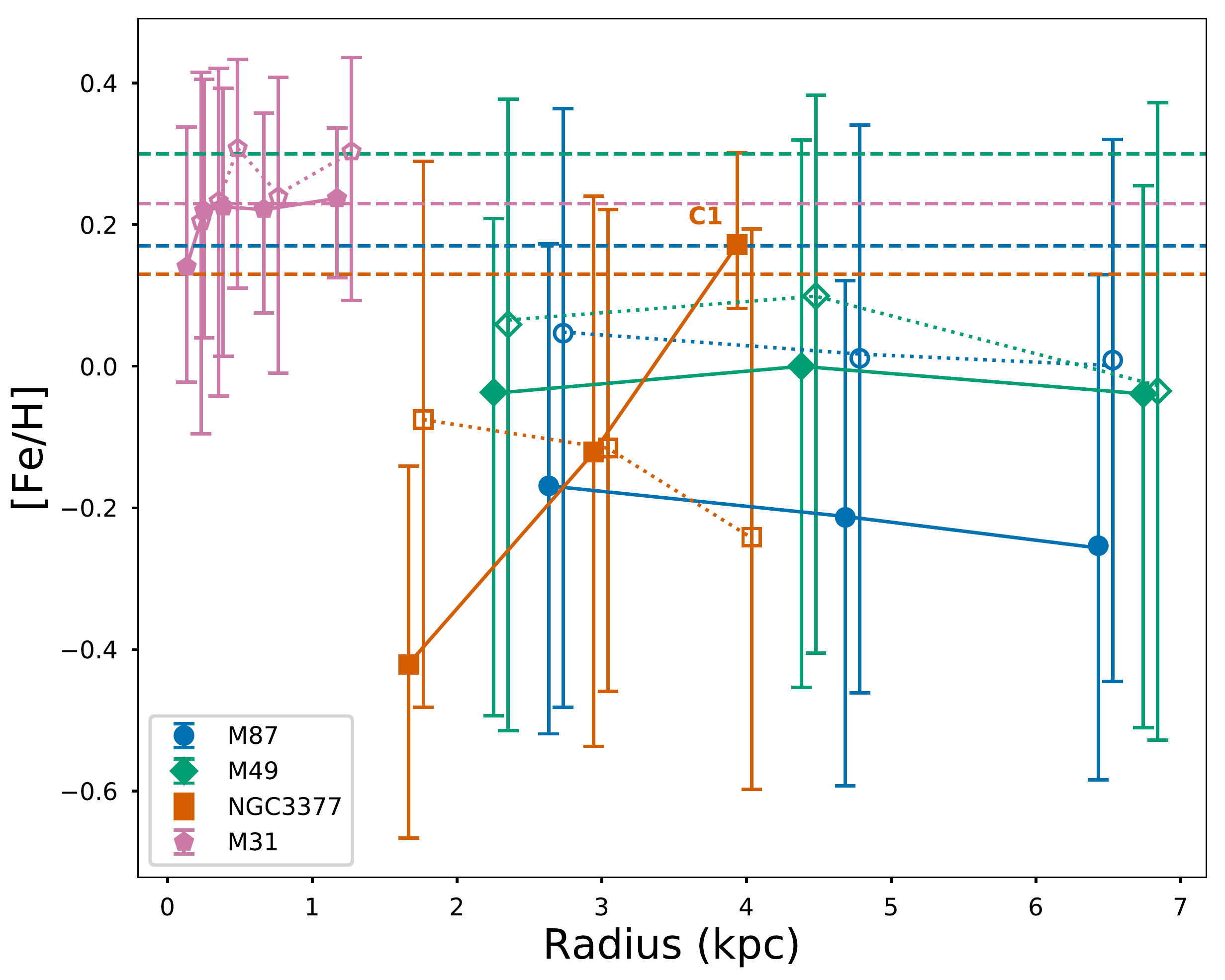}
\caption{Measurements of average \FeH{} as a function of radius. Symbols as in \F{f.pcmdpy2.younggrads}.
There is no evidence for abundance gradients, within the large uncertainties. Region C1 of NGC 3377 is noted, which had a poorly-fit distance.
\changed{The horizontal lines indicate the literature \ZH{} measurements from \T{t.pcmdpy2.distances}.}}
\label{f.pcmdpy2.feh_grads}
\end{figure}

The \pcmdpy{} models also derive estimates of the metal enrichment, in terms of the iron-abundance \FeH{}.
\F{f.pcmdpy2.feh_grads} shows the metallicity derived at each radius in the galaxies studied, although we emphasize that the metallicity is at best poorly constrained in most regions. This is primarily due to the strong degeneracy between dust and metallicity (see Figures \ref{f.pcmdpy2.m31corner} and \ref{f.pcmdpy2.m87corner}), which the models are not able to easily distinguish between at this high \Npix{}. Given the large uncertainties on measured metallicity, we can place no significant constraints on abundance gradients at the radii studied.

\changed{The figure also includes the literature \ZH metallicities given in \T{t.pcmdpy2.distances}, which are not spatially resolved.
The measured metallicities in M31 agree very well with the literature.
The metallicities measured in the other three galaxies are generally too low, although they are within the large uncertainties.
In models that assume a closed-box MDF, the disagreements are mostly less than 0.2 dex. 
Better precision metallicity estimates from high \Npix{} pCMDs should be possible with improvements to the likelihood model (see \S{ss.pcmdpy2.likelihood}).}

It is important to note that the MIST models \citep{Choi2016} used in \pcmdpy{} assume a solar $\alpha$-element ratio, whereas many massive elliptical galaxies (such as M87 and M49) show strong evidence for super-solar $\alpha$-abundances and $\alpha$-abundance gradients \citeeg{Sarzi2018}.
\changed{Therefore, the literature metallicities, which include variable abundances, are not perfectly comparable with our measurements.}

$\alpha$-enrichment has been shown to affect broadband colors in the RGB and MS in old stellar populations \citep{Dotter2008}, producing redder isochrones.
This effect is likely relatively minor, and the offsets could have been absorbed in the model fits by the reddening contribution from dust.
Because of this, we do not study the measurements of dust in close detail.

\section{Discussion: Caveats and Future Applications}\label{s.pcmdpy2.discussion}

\subsection{Likelihood Models for pCMDs}\label{ss.pcmdpy2.likelihood}

We believe one of the most significant remaining limitations to better constrained distances and stellar populations, as well as applying the pCMD technique to more datasets and with more complex models, is identifying a better likelihood model for comparing observed and simulated pCMDs.
Comparing distributions of two (or more) dimensional datapoints is a challenging and unsolved problem, yet this is necessary for comparing data and model pCMDs.

In \citet{Cook2019}, we describe our choice to approximate the likelihood through comparing the number of simulated and data pixels in bins in Hess diagram space, using the size of the bins as a rough control for the expected measurement uncertainty.
However, regardless of whether we assume a Gaussian likelihood model, a Poisson likelihood, or something else (we have experimented with many alternative methods), we are left with the question of whether the likelihood derived in this manner is statistically meaningful.
In other words: can we reasonably say that one set of parameters are ten times more likely to have led to the data if this likelihood metric is ten times higher than for another set?

Even leaving aside stochastic effects (discussed in detail in Appendix A of \citet{Cook2019}), we are not convinced the likelihood framework used here is ideal.
Quite often, a very small change in input parameters (such as changing $\FeH$ by 0.02) can result in extreme changes ($\gg 10^3 \times$) to the likelihood.
We find this over-specification effect is most significant where the data and model are very similar.
The likelihood model adopted here appears reasonable for discriminating between poor and decent fits, but is oversensitive to small changes to already reasonable fits.
This over specification is primarily due to the edge effects introduced by binning, and we suggest that future work be dedicated to identifying alternative metrics that do not rely on binning the pCMDs.

This behavior led us to the likelihood adjustment procedure described in \S{ss.pcmdpy2.convergence}, which significantly down-weights the samples with the highest likelihood.
Philosophically, this should be reasonable as we believe that we cannot quantitatively trust the relative likelihoods in the very best-fit points.
We believe the samples with the highest likelihoods are probably somewhat better fits, but we do not trust the quantitative values.
This correction procedure will almost certainly result in overestimating the uncertainties. 

A more statistically appropriate likelihood model would go a long way towards improving the precision with which \pcmdpy{} can constrain distances and stellar populations.
One possible way forward we have considered is to use a 2-D version of the Kolmogorov-Smirnov (KS) test to evaluate likelihoods.
The KS test is a well established metric for comparing fairly arbitrary distributions of points in one dimension by comparing the cumulative distribution functions.
But the KS test does not generalize easily to two or more dimensions, because there is no principled way to order points in a cumulative fashion.
We have experimented with using the structure of the pCMDs themselves to define an ordering, by creating a dense series of smoothed contours around the data pCMD points (like those shown in Figures \ref{f.pcmdpy2.M87pcmds} and \ref{f.pcmdpy2.residuals}), and comparing the relative number of model points that fall within each contour.
An alternative likelihood model based on this concept is currently in development for \pcmdpy{}.

\subsection{Applying the pCMD Method at Larger Distances}\label{ss.pcmdpy2.ngc4993}

In this work, we have focused on elliptical galaxies within 20 Mpc.
As the distance to a galaxy increases, the observed \Npix{} will correspondingly increase, because the same number of stars will be confined to fewer pixels. 
The magnitude of surface-brightness fluctuations scales roughly as $\Npix^{-\frac{1}{2}}$, due to the Poisson sampling of the rare stars.
While this implies the precision with which SFHs and abundances can be inferred will decline at larger distances, in principle there is no reason why they should be completely unmeasurable.

To demonstrate this, we also attempted to fit \pcmdpy{} models to archival photometry of NGC 4993, the elliptical galaxy at approximately 40 Mpc that was identified as the host galaxy of GW170817, the first binary neutron star merger detected in gravitational waves \citep{Cantiello2018}. 
The data could not be well fit by \pcmdpy{}, as the best-fit models predict a distance of less than 10 Mpc, more than $4\times$ closer than the SBF-measured distance.

Despite the failure in our only galaxy studied well beyond 20 Mpc, we believe that extending application of the pCMD technique to galaxies as far as 50 Mpc should be possible. The primary practical limitations to extending to larger distances are twofold, and both play a role in the failure of fitting NGC 4993.

First, at larger distances, galaxies necessarily extend a smaller area on the sky, reducing the number of adjacent pixels over which the constant \Npix{} approximation is valid.
In NGC 4993, not only is the radial profile a concern, but we found evidence of notable ($\sim20\%$) azimuthal variations in the mean magnitudes of the regions we extracted, especially in the inner, $\sim3$ kpc, annulus.
This is roughly equivalent to averaging over several \Npix{} values, and has the effect of increasing the dispersion in the pCMD.
\citet{Blanchard2017} previously detected this extended substructure, including radial shells and strong azimuthal asymmetries, which they argue is evidence of an active history of recent mergers.

Secondly, significantly longer exposure times may be necessary to model pCMDs at large distances (and therefore higher \Npix).
Because the average flux per pixel remains constant with distance, so will the error contribution from photon noise, such that at fixed exposure time there will necessarily come a distance beyond which any intrinsic fluctuations in surface-brightness will be drowned out by observational errors \citep[see Figure 7 of][]{Cook2019}.
In NGC 4993, given the large distance and relatively short exposure time of the archival photometry, the flux from the galaxy was as little as $50\%$ of the estimated sky background in the outer region, and still only $150\%$ above the background in the inner region.
Any uncertainty in the background estimation is therefore extremely important, and likely also contributed to the poor fits.

Additional care in the data reduction stage must therefore be taken when attempting to apply the pCMD technique to more distant galaxies.
Highly precise measurements of the sky noise will be essential, as will longer exposure times, or stacking of many overlapping exposures.
Yet, as mentioned in \S{s.pcmdpy2.data}, stacking many exposures also requires careful consideration of which pixels were contributed to by all exposures, in order to accurately model the properties of photon noise in the simulations.
Nonetheless, these are all issues that can be addressed with careful attention to the details of the photometry and proper implementation in the modeling process.

\subsection{Point-Spread Function Models}\label{ss.pcmdpy2.psf}

In this work, we assume all stars reside at the centers of their respective pixels, while in practice the positions of stars are randomly distributed within a pixel.
This effect can have a noticeable impact on the distribution of pCMDs, as a bright star near the edge of a pixel will deposit more of its light in neighboring pixels than one at the center, changing the distribution of the surface-brightness fluctuations.

The ideal method to replicate this effect would be to subsample the simulated pixels (we will call this the \textit{subsampled PSF} method). In this approach, stars are populated (for instance) into a $3\times3$ grid representing sub-pixel positions, their light is distributed using a sub-pixel PSF model, and then accumulated into the resulting pixels.
This approach adds somewhat to the computational cost of simulating a pCMD (effectively increasing $\Nim$ by a factor of $3$ in this example) but would be fairly straightforward.

The primary limitation to this approach is access to detailed sub-pixel models for the \HST{}-ACS PSF, as the latest verison of \texttt{Tiny Tim} does not provide sub-pixel PSF models appropriate for drizzled \HST{} photometry. The raw (un-drizzled) photometry is of insufficient quality to perform pCMD analysis due to geometrical distortions, low individual exposure times, and cosmic ray contamination.

An alternative approach outlined in \citet{Conroy2016} and \citet{Cook2019}, which we will call the \textit{dithered PSF} method, attempts to replicate sub-pixel effects by applying a grid of PSFs, shifted by fractions of a pixel, to different regions of the simulated image.
This approach therefore assumes all stars in one section of an image reside at the center of a pixel, while all stars in another section reside near the edges.

To test the applicability of the dithered PSF assumption, we compared it to the "correct" subsampled PSF by simulating pCMDs with each technique and assuming a Gaussian PSF, so we can easily generate a sub-pixel PSF model.
The dithered PSF approach was found to artificially decrease the variation in simulated pCMDs, by smoothing the light of stars in some regions of the image over too many pixels.
By contrast, pCMDs generated using a single PSF across the entire image (assuming all stars reside at the centers of their pixels) were remarkably similar to those assuming a subsampled PSF, although distinct artifacts in pCMD space appear when $\Npix \lsim 10^3$.
We therefore ignored sub-pixel effects in this work, but emphasize that extending the pCMD technique to lower \Npix{} (such as in the disk of M31) will require a detailed sub-pixel PSF model for the \HST{}-ACS camera, applicable to drizzled photometry.

\subsection{Beyond Intrinsically Smooth Populations}\label{ss.pcmdpy2.moregals}

In addition to extending the pCMD technique to larger distances, there remains the potential of applying it to other nearby systems like spiral galaxies. 
This work focused on massive elliptical galaxies because of their relative simplicity.
The complex structures in disk galaxies present more significant challenges to the simplifying assumptions that make fitting \pcmdpy{} models tractable\changed{: namely, that the stars (and dust) in each pixel represent independent draws from the same underlying statistical distribution with mean number $\Npix$.
These assumptions are much poorer approximations in late-type galaxies, which have complex features like spiral arms, dust lanes, and star forming clusters.}

\changed{We see two potential approaches that could be used to apply pCMDs to late-type galaxies. The first would involve adding more levels of complexity to the forward modeling procedure. For instance, each simulated image could be represented as having multiple regions with different \Npix{} (perhaps representing star clusters, spiral arms, or arm gaps) and different dust properties (representing thick dust lanes).}
Adding additional model complexity is relatively straightforward: we have already experimented with allowing a distribution of \Npix{} or a two-component dust model, \changed{and the overhead of creating more complex simulated images can be minimized thanks to the GPU-acceleration used in \pcmdpy{}. But adding these additional components and parameters, many of which are highly degenerate, makes it intractable to fit such complex models to data.}

A better path forward to studying the pCMDs of late-type galaxies may be through more sophisticated data reduction methods\changed{, with the goal of extracting relatively uniform sub-regions, within which the assumptions made by \pcmdpy{} are more valid. For instance, the assumption of constant \Npix{} might be a reasonable approximation of the distribution of stars within a spiral arm gap, or within a small region of a spiral arm (if dust lanes and star clusters are isolated out).}
Just as we apply \pcmdpy{} to elliptical regions in the galaxies above, disk galaxies could be partitioned \changed{into these sub-regions,} perhaps through something similar to Voronoi tessellation \citeeg{Cappellari2003}.

\subsection{Complementarity of the PCMD Technique with Existing Methods}\label{ss.pcmdpy2.young}

In \S{ss.pcmdpy2.SFHs}, we demonstrated the capability of the pCMD method to constrain the relative amounts of old ($\gsim 2$ Gyr) star formation in most of the systems studied.
The fits were only capable of measuring upper limits on the younger epochs of star formation, although in a few cases (primarily in M31) the constraints are sufficient to indicate a significant decrease in the star-formation rate in the most recent $\sim 1$ Gyr.
Yet we believe it to be a fairly generalizeable result that pCMDs, in the current framework, are limited in their capability of measuring low levels of young star formation.

The primary reason for this is simply the rarity of young stars, on a pixel-by-pixel basis.
While massive, young stars are extremely bright and blue, and will therefore dominate the flux of any pixels they populate, they will contribute to only a very small fraction of the pixels in most pCMDs.
Even in the case of a fairly young ($\tau \sim 5$ Gyr) population, the cumulative contribution of all stars younger than 10 Myr is around 1 part in $10^4$.
This requires a relatively dense ($\Npix > 10^4$) region for there to be one young star per pixel, on average, and even then the majority will be low-mass, main sequence stars, indistinguishable photometrically from their older counterparts.
Therefore, only a handful of rare pixels will be contributed to by young, massive stars, posing a challenge for our current likelihood models to constrain the amount of young star formation in all but the densest or youngest systems.

The pCMD technique, which can constrain old histories of star formation, ought to therefore be highly complementary to the resolved star and integrated light methods, which are useful for constraining young star formation but often insensitive to the oldest ages.
Unfortunately this complementarity, both in the systems and epochs of star formation that can be studied, makes it challenging to validate the SFHs measured with the pCMD technique.
As first shown in \citet{Conroy2016}, pCMDs can measure old SFHs in the dense bulge of M31, where the resolved star technique fails to resolve the oldest main sequence turnoff due to crowding.
The entire history of star formation should therefore be recoverable by combining the resolved star and pCMD methods: bright, young stars can be first be resolved and photometered and the remaining crowded pixels used to study the old star formation with \pcmdpy{}.

Integrated light methods are also more sensitive to young star formation because they sum the flux over much larger regions than individual \HST{}-ACS pixels, therefore including more of the rare, young stars.
But these very bright stars typically overwhelm the faint background of old main sequence stars \citep[the "outshining" effect,][]{Maraston2010,Pforr2012a,Sorba2015}, and the ancient SFH is therefore often degenerate with other sources of red light, such as dust or AGN activity.
The constraints placed on the SFH by the shape of the assumed prior distribution places additional challenges on the interpretations \citep{Carnall2018b,Leja2019}.
\changed{For instance, our measurements of the old SFH in M87 and M49 indicate that the young and old histories of SFHs may not be well represented with a single $\tau$ model.
If that is true, then integrated light methods, which are highly sensitive to the most recent star formation, may be underestimating the amount of old SFH when assuming a $\tau$ prior.}
Where high-resolution \HST{} imaging overlaps with multi-band photometry or spectroscopy, combining the pCMD and SED-modeling techniques could therefore be useful in constraining the entire SFH.

\section{Summary}\label{s.pcmdpy2.conclusion}

In this work, we present the first application of the pixel color-magnitude diagram (pCMD) technique to galaxies outside the Local Group. 
We construct spatially-resolved pCMDs from archival \HST{} photometry in three nearby elliptical galaxies (M87, M49, NGC 3377) and the M31 bulge.
We align and clean the archival data, and extract the pCMDS from thin elliptical regions aligned with the galaxy's orientation, so that the properties of the data match the underlying pCMD modeling assumptions (especially that of fixed \Npix{}) as closely as possible. We use the new GPU-accelerated code \pcmdpy{} to fit each dataset to model pCMDs, and derive Bayesian posteriors over a 5-bin non-parametric star formation history, distance modulus, and metallicity.

We summarize the main results as follows:

\begin{enumerate}
    \item
    We derive spatially-resolved SFHs in each of the four galaxies.
    In the nearest galaxies, M31 and NGC 3377, we constrain the relatively young ($\lsim 2$ Gyr) history of star formation, and find notable evidence for radial gradients in the SFH.
    The stars in the inner regions formed significantly earlier than in the outer regions.
    In M49 and M87, only the very oldest ages of star formation are constrained.
    
    \item
    At similar radii, the bulge of M31 has a younger SFH than NGC 3377, which in turn shows evidence of more recent star formation than M87 or M49.

    \item
    We recover distance estimates to the four galaxies, in good ($\sim 10\%$) agreement with the tip of the red-giant branch (TRGB) and surface-brightness fluctuation (SBF) techniques through combining the individual distance estimates in each region.

    \item 
    pCMDs are more sensitive to the oldest epochs of star formation, and so the pCMD technique should be highly complementary to existing SFH measurement techniques that primarily constrain young star formation.

\end{enumerate}

\acknowledgments
We thank Peter Blanchard and Ashley Villar for helpful discussions about the details of \HST{} photometry and data reduction. B.C.~acknowledges support from the NSF Graduate Research Fellowship Program under grant DGE-1144152. This work is supported in part by HST-AR-14557. The computations in this paper were run on the Odyssey cluster supported by the FAS Division of Science, Research Computing Group at Harvard University.

\software{This research has made use of NASA's Astrophysics Data System, as well as the following software packages: PyCUDA \citep{Klockner2012}, Dynesty \cite{Speagle2019}, NumPy \citep{VanderWalt2011}, Matplotlib \citep{Hunter2007}, IPython \citep{Perez2007}, Jupyter \citep{Kluyver2016}, SciPy \citep{Jones2001},
Pandas \citep{McKinney2010}, and Astropy \citep{TheAstropyCollaboration2013,TheAstropyCollaboration2018}.}

\bibliography{references}
\end{document}